\documentclass[%
 aip,
 amsmath,amssymb,
 reprint,%
]{revtex4-1}

\usepackage{graphicx}
\usepackage{dcolumn}
\usepackage{bm}
\usepackage{hyperref}
\usepackage{color}
\usepackage{epsfig}
\usepackage{epstopdf}
\usepackage{graphicx}

\usepackage[utf8]{inputenc}
\usepackage[T1]{fontenc}
\usepackage{mathptmx}

\begin{document}

\preprint{AIP/123-JAP}

\title[]{Electronic Structure and Thermoelectric Properties of Half-Heusler Alloys NiTZ}

\author{Dhurba R. Jaishi}
 \affiliation{Central Department of Physics, Tribhuvan University, Kirtipur 44613, Kathmandu, Nepal}
 \affiliation{Condensed Matter Physics Research Center (CMPRC), Butwal 32907, Rupandehi, Nepal}
\author{Nileema Sharma}
\affiliation{Central Department of Physics, Tribhuvan University, Kirtipur 44613, Kathmandu, Nepal}
\affiliation{Condensed Matter Physics Research Center (CMPRC), Butwal 32907, Rupandehi, Nepal}
\author{Bishnu Karki}
\affiliation{Central Department of Physics, Tribhuvan University, Kirtipur 44613, Kathmandu, Nepal}
\affiliation{Condensed Matter Physics Research Center (CMPRC), Butwal 32907, Rupandehi, Nepal}

\author{Bishnu P. Belbase}
\affiliation{Central Department of Physics, Tribhuvan University, Kirtipur 44613, Kathmandu, Nepal}
\affiliation{Condensed Matter Physics Research Center (CMPRC), Butwal 32907, Rupandehi, Nepal}

\author{Rajendra P. Adhikari}
\affiliation{Department of Physics, Kathmandu University, Dhulikhel 45200, Nepal}

\author{Madhav P. Ghimire}
\email{madhav.ghimire@cdp.tu.edu.np}
\affiliation{Central Department of Physics, Tribhuvan University, Kirtipur 44613, Kathmandu, Nepal}
\affiliation{Condensed Matter Physics Research Center (CMPRC), Butwal 32907, Rupandehi, Nepal}

\date{\today}

\begin{abstract}
We have investigated the electronic and thermoelectric properties of half-Heusler alloys NiTZ (T = Sc, and Ti; Z = P, As, Sn, and Sb) having 18 valence electron. Calculations are performed by means of density functional theory and Boltzmann transport equation with constant relaxation time approximation, validated by NiTiSn. The chosen half-Heuslers are found to be an indirect band gap semiconductor, and the lattice thermal conductivity is comparable with the state-of-the-art thermoelectric materials. The estimated power factor for NiScP, NiScAs, and NiScSb reveals that their thermoelectric performance can be enhanced by appropriate doping rate. The value of $ZT$ found for NiScP, NiScAs, and NiScSb are 0.46, 0.35, and 0.29, respectively at 1200 K. 
\end{abstract}

\maketitle

%

 \section{Introduction}
In the past few decades, researchers have been focused on the investigation of the multi-functional materials, which can be used as various applications such as in spintronics, optoelectronics, thermoelectrics (TE), etc. With the surge in demand for green energy sources, TE materials are extensively taken into considerations for their ability to convert relatively small and waste heat into useful energy at the time of energy consumption. Wide range of materials has been explored for the potential half-metals and TE devices such as organic \cite{russ2016organic}, chalcogenides \cite{kanatzidis2010nanostructured,snyder2011complex}, skutterudites \cite{lan2010enhancement,szczech2011enhancement,shankar2017electronic}, oxides \cite{ghimire2010study,ghimire2015half,ghimire2016compensated,roy2016environmentally,
choi2015polaron,bhandari2020electronic}, hybrid perovskites\cite{filippetti2016appealing,lee2015organic,liu2017extremely}, triple-point metals \cite{PhysRevMaterials.2.114204}, ternary compounds\cite{rai2017electronic}, and half-Heusler (hH) alloys \cite{kim2007high,fu2015realizing,zhu2018discovery, zhu2019discovery,graf2010phase,PhysRevB.83.085204,zeeshan2018fetasb,singh2019first,
PhysRevMaterials.1.075407,PhysRevMaterials.1.074401,singh2019textit,rai2015study}. Among them, Heusler compounds have gained much more attention since their discovery in 1903 due to their simple crystalline structure with fascinating properties that includes magnetism, half metallicity, superconductivity, optoelecronic, piezoelectric semiconductors, thermoelectricity, topological insulators and semimetals \cite{sakurada2005effect, PhysRevLett.109.037602, PhysRevB.69.134415, PhysRevB.82.235121, PhysRevLett.50.2024, ghimire2011first, ghimire2012magnetic, nakajima2015topological,zhang2020topological}. 

Thermoelectric materials are found applicable in day-to-day lives to fulfill the increasing demand of energy of the globalized society. The highly efficient TE devices (cooler, power generator, temperature sensors, etc) can utilize a large amount of wasted thermal energy to generate electricity and vice-versa \cite{zaitsev2006thermoelectrics,riffat2003thermoelectrics}. For this, the device needs a larger figure of merit ($ZT$), which depends on the transport properties \cite{pei2011convergence,lalonde2011lead} defined by
\begin{equation}
ZT=\frac{\alpha^2\sigma T}{\kappa}
\label{eq1}
\end{equation}
where $\alpha$ (V K$^{-1})$ is the Seebeck coefficient, $\sigma$ (S m$^{-1})$ is the electrical conductivity, $\kappa$=$\kappa_e$+$\kappa_l$ (W m$^{-1}$K$^{-1}$) is thermal conductivity, and T(K) is the absolute temperature. 
$\alpha$$^{2}$$\sigma$ is defined as the power factor (PF). The symbol $\kappa_e$ and $\kappa_l$ represents the electronic and lattice thermal conductivity, respectively. The materials having a high value of PF along with the low value of $\kappa$ are suitable for the efficient TE devices \cite{sootsman2009new}.

Among others, most of the cubic hH alloys with 18 valence electron count (VEC) exhibits high Seebeck coefficients and are reported as promising materials for TE application due to high electrical conductivity and narrow band gap semiconductors with novel electrical and mechanical properties even at high-temperatures \cite{fu2015realizing,zhu2018discovery, zhu2019discovery,graf2010phase,felser2007spintronics}. In addition to it, hH alloys contain non-toxic and readily available elements, making them environmentally friendly and more cost effective. 

Recent experimental and theoretical investigations on hH alloys are mainly focused on improving their thermoelectric efficiency $ZT$ by tuning the power factor and thermal conductivity. Band gap engineering and fluctuation of carrier concentration around the Fermi level ($E_{\rm F}$) in $Z$ position is a widely used method to enhance the power factor, whereas, thermal conductivity can be decreased by alloying or by doping on $X$ or $Y$ site to fluctuate the mass of the carriers introducing impurities and nanostructuring \cite{heremans2008enhancement,zhao2014ultralow,biswas2012high,toberer2011phonon}.

From the literatures, we noticed that Ni-based hH alloys with 18-VEC are less investigated. Following the Slater-Pauling's rule, the total magnetic moment for these type of hH alloys should be zero. Thus, the zero moment on Ti or Sc at $Y$ site and P or As or Sb at $Z$ site will give rise to zero moment for the Ni atom at the $X$ site resulting in a non-magnetic system\cite{ma2017computational, kandpal2007calculated}. This motivates us to explore the electronic, TE, and other related properties to confirm if these groups of materials could be suitable for TE devices.

\section{Computational Details}
\begin{figure}[h!]
	\centering
	\includegraphics[width=2.5in,height=2in]{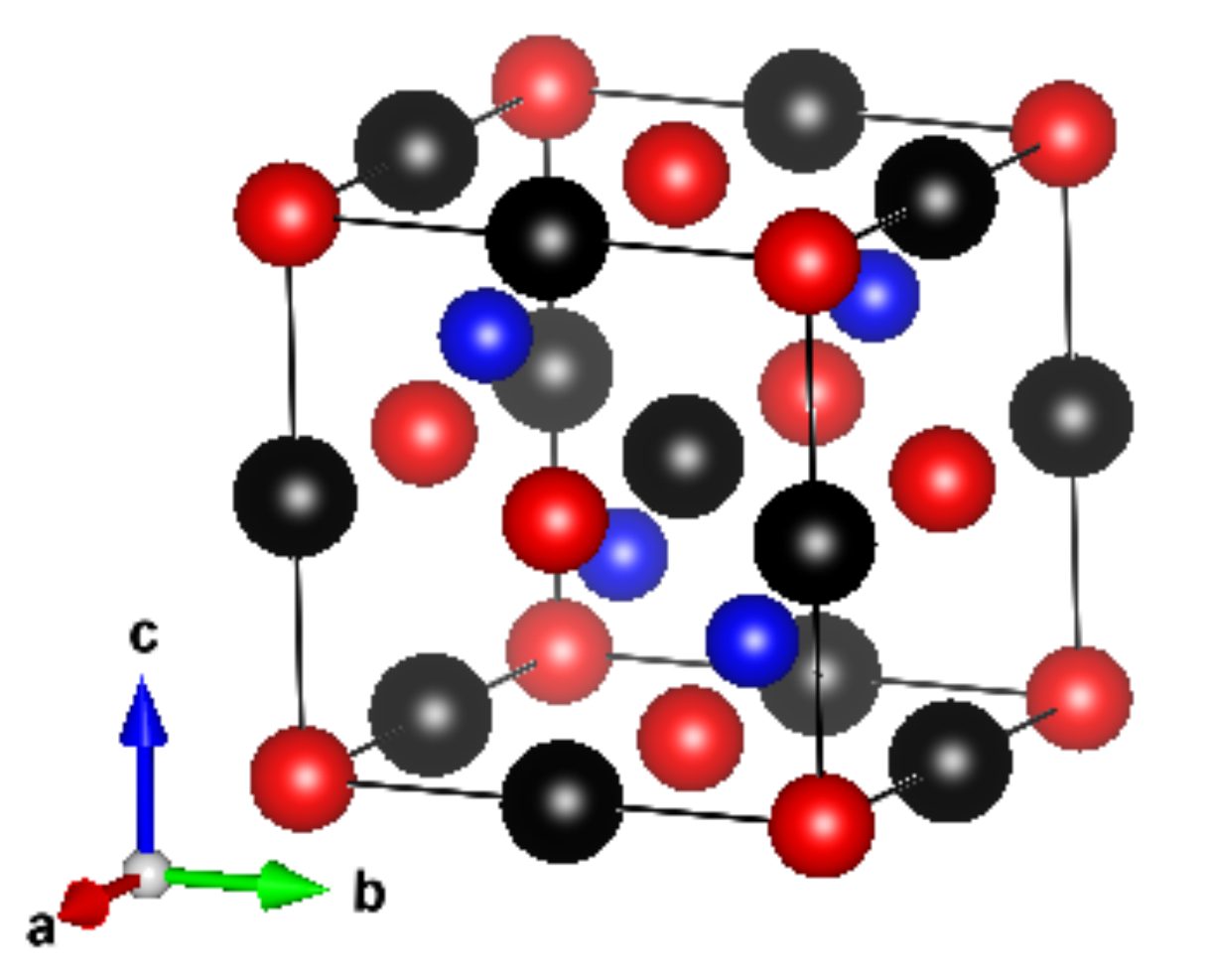}
	\caption{The crystal structure of cubic hH NiTZ (T= Sc, and Ti; Z= P, As, Sn, and Sb). The balls in blue,  black, and red color represents Ni, T, and Z atoms, respectively.}
	\label{Fig1}
\end{figure}

The cubic hH alloys NiTZ (T= Sc, and Ti; Z= P, As, Sn, and Sb) belongs to Cl$_b$ structure with space group $F\bar{4}3m$. It contains three in-equivalent atoms forming inter-penetrating fcc sublattices with the Wyckoff positions Ni (1/4, 1/4, 1/4), T (1/2, 1/2, 1/2) and Z (0, 0, 0), respectively as shown in Figure \ref{Fig1}. The iso-structural NiTiSn is used here to validate our calculations based on the earlier reported results (both theoretical and experimental).

The density functional (DF) calculations has been performed using the full-potential linear augmented plane wave (FP-LAPW) method as implemented in the WIEN2k code \cite{blaha2001wien2k}. We double checked some parts of our calculations using the plane-wave based pseudopotentials Quantum Espresso (QE) package \cite{giannozzi2009quantum}. The standard generalized-gradient approximation (GGA) in the parameterization of Perdew, Burke, and Ernzerhof (PBE) \cite{perdew1996generalized} was used in scalar-relativistic mode. The modified Becke-Johnson (mBJ) potential \cite{PhysRevLett.102.226401} was further included to check the accuracy of the band gaps. The self-consistency convergence criteria for charge was set to 10$^{-4}e$, and energy to 10$^{-5}$ Ry. 

In the plane-wave pseudopotential approach, we used the norm-conserving pseudopotentials with plane wave cut-off energy for wave function set to 90 Ry. The full Brillouin Zone (BZ) was sampled with an optimized 10$\times$10$\times$10 mesh of Monkhorst-Pack $k-$points. To check the dynamical stability, phonon spectrum calculations have been performed with 4$\times$4$\times$4 $q-$ mesh in phonon BZ, which is based on the DF perturbation theory (DFPT) implemented in the QE package \cite{giannozzi2009quantum}. 

The TE properties were calculated using the Boltzmann semi-classical transport equation and constant relaxation time approximation based on a smoothed Fourier interpolation of the bands implemented on BoltzTraP code \cite{madsen2006boltztrap}. The full BZ was sampled with $50\times50\times50$ $k-$ mesh for the calculation of the transport properties. The electrical conductivity and PF were calculated under constant relaxation time approximation ($\tau$) using the BoltzTraP code based on Boltzmann theory. $\tau$ is approximated by fitting the experimental data from Kim $et$ $al$. \cite{kim2007high}. The lattice thermal conductivity was obtained by solving linearized Boltzmann transport equation (BTE) within the single-mode relaxation time approximation (SMA) using thermal2 code implemented in QE package\cite{giannozzi2009quantum}.

  \section{Results and Discussion}
\subsection{Structure Optimization and Phonon Stability}
We started our calculations by optimizing the cubic hH  alloys with $F\bar{4}3m$ symmetry. Our calculated values of lattice parameters and the band gap within GGA and GGA + mBJ are listed in Table \ref{tab1}. These values are found to be in fair agreement with the earlier reports of Ma $et$ $al.$ \cite{ma2017computational} for the GGA case. 

\begin{table}[h!]
	\caption{\label{tabone}The optimized lattice constant $a$ and the band gap $E_{g}$ within GGA and GGA + mBJ for the cubic hH alloys NiTZ.} 
	\centering
\begin{tabular}{ccccccccccccc}
	\hline\hline
	&&&&&&&GGA&&GGA+mBJ\\
	\hline
	System & & & & a (\AA) & & & E$_g$(eV) & &E$_g$(eV)& \\
	\hline\\
	NiScP  & & & & 5.69 & & & 0.54& & 0.62 \\ \\
	NiScAs & & & & 5.84 & & & 0.48& & 0.52 \\ \\
	NiScSb & & & & 6.12 & & & 0.28& & 0.32 \\ \\
	NiTiSn & & & & 5.95 & & & 0.46& & 0.45 \\
	\hline\hline
\end{tabular}
\label{tab1}
\end{table}

\begin{figure}[h!]
	\centering
	\includegraphics[width=\columnwidth,height=3.5in]{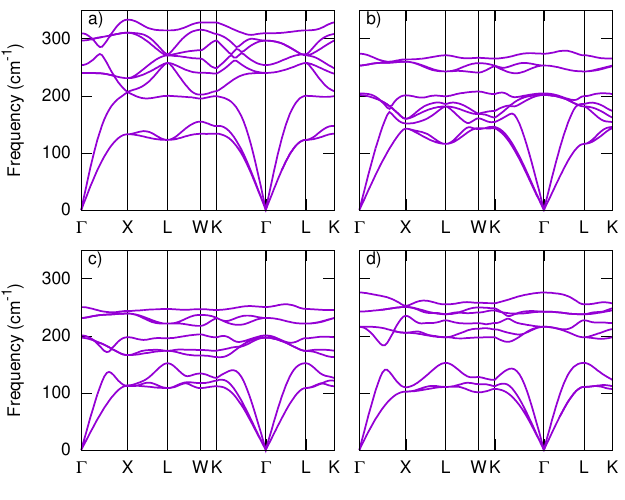}
	\caption{Phonon band structures for finding the dynamic stability of a) NiScP, b) NiScAs, c) NiScSb, and d) NiTiSn.}
	\label{Fig2}
\end{figure}

The calculated phonon dispersion curves along the high-symmetry points shown in Figure \ref{Fig2} depicts that the proposed hH alloys are thermally stable. This is evidenced by the absence of imaginary phonon frequencies throughout the whole BZ, as expected for dynamic stability \cite{togo2015first}. We observed three acoustic (low-frequency region) and six optical phonon (high-frequency region) branches due to three atoms per unit cell. The majority of the lattice contribution to the thermal conductivity arises from the acoustic part as it has high group velocity compared to the optical part. We found that the acoustical phonon branches of NiScP and NiScAs extends nearly to 200 cm$^{-1}$ while NiScSb and NiTiSn lies within 150 cm$^{-1}$ in frequency. 
The observation of dynamical stability and preferable energy gap in our proposed hH alloys motivate us to explore the electronic and transport properties for their potential application as TE materials.

\subsection{Electronic Properties}
\begin{figure}[h!]
	\centering
	\includegraphics[width=\columnwidth, height=3.5in]{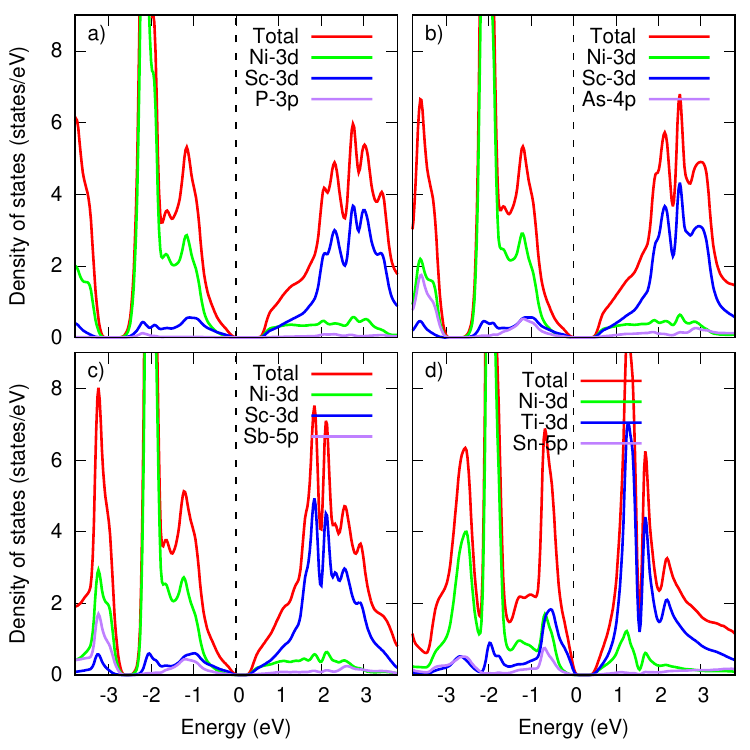}
	\caption{Total and Partial density of states of a) NiScP, b) NiScAs, c) NiScSb, and d) NiTiSn within GGA + mBJ. Vertical dotted line represent $E_{\rm F}$.}
	\label{Fig3}
\end{figure}

To understand the ground state electronic properties of the material, the total and partial density of states (DOS) are shown in Figure \ref{Fig3}. The proposed systems are found to be semiconducting with an energy gap lying within $\sim 0.32-0.62$ eV, in fair agreement with the earlier report \cite{ma2017computational}. 
As seen in the PDOS the main contribution to the total DOS at and around $E_{\rm F}$ are from the $3d$-orbitals of Ni and Sc atoms while the contributions from atom on the Z site is negligible (see  in Figure \ref{Fig3}). This is an indication that doping onto the Z site may improve the carrier concentration. .

\begin{figure}[h!]
	\centering
	\includegraphics[width=1.7in,height=2in]{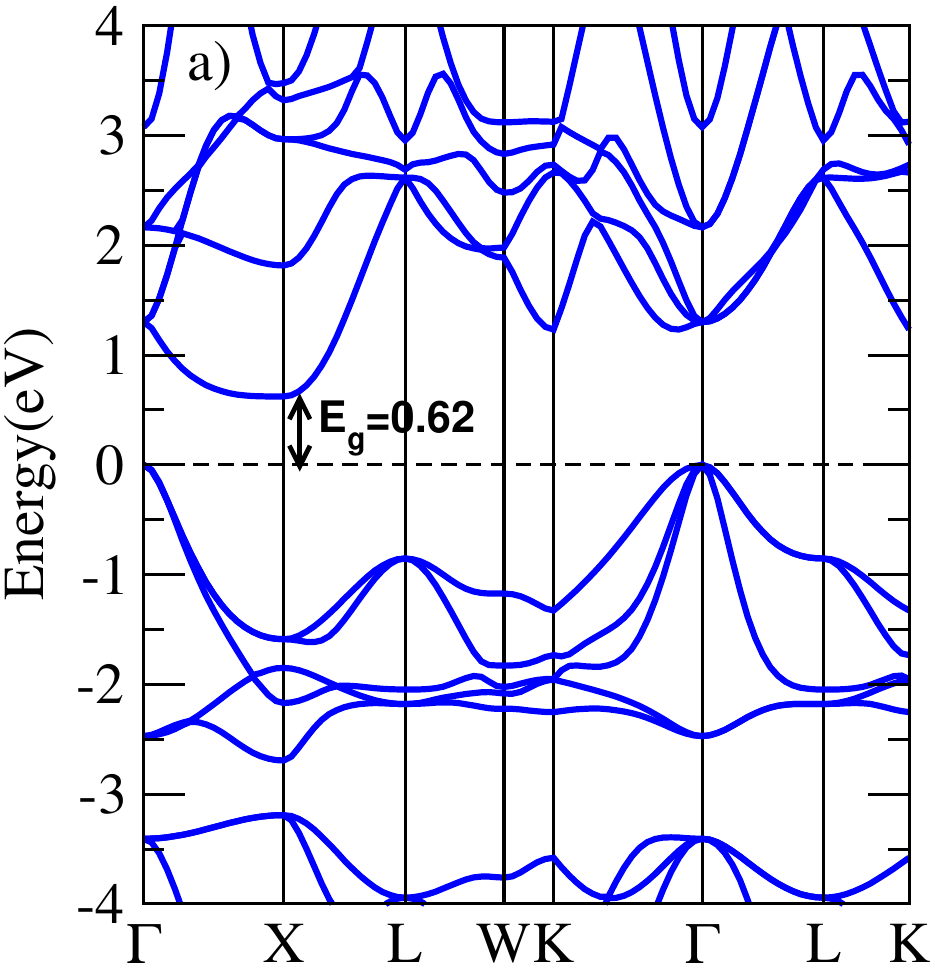}%
	\includegraphics[width=1.7in,height=2in]{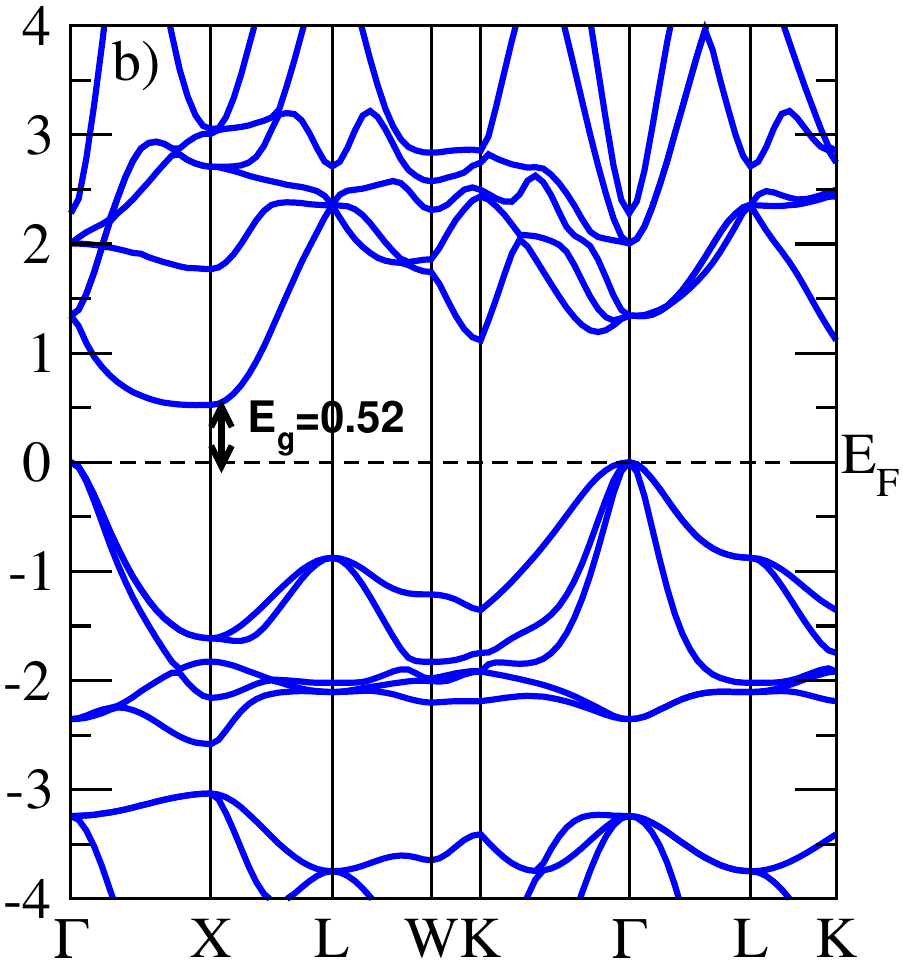}
	\includegraphics[width=1.7in,height=2in]{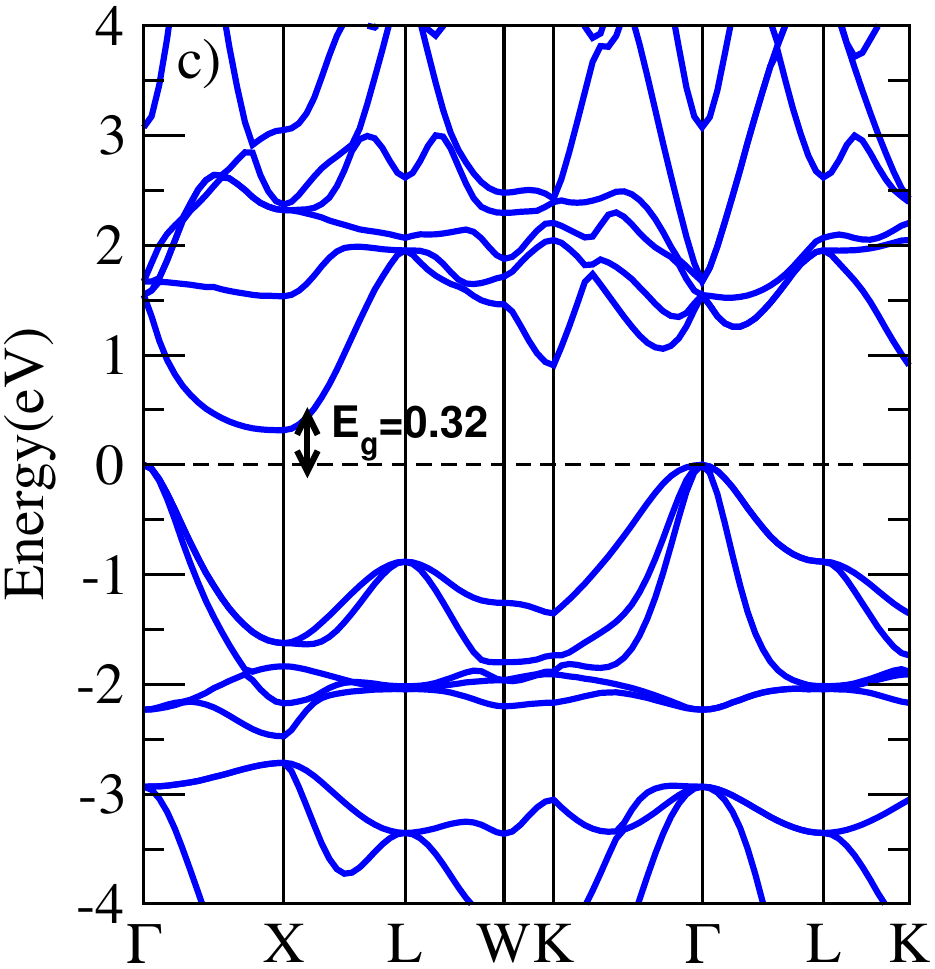}%
	\includegraphics[width=1.7in,height=2in]{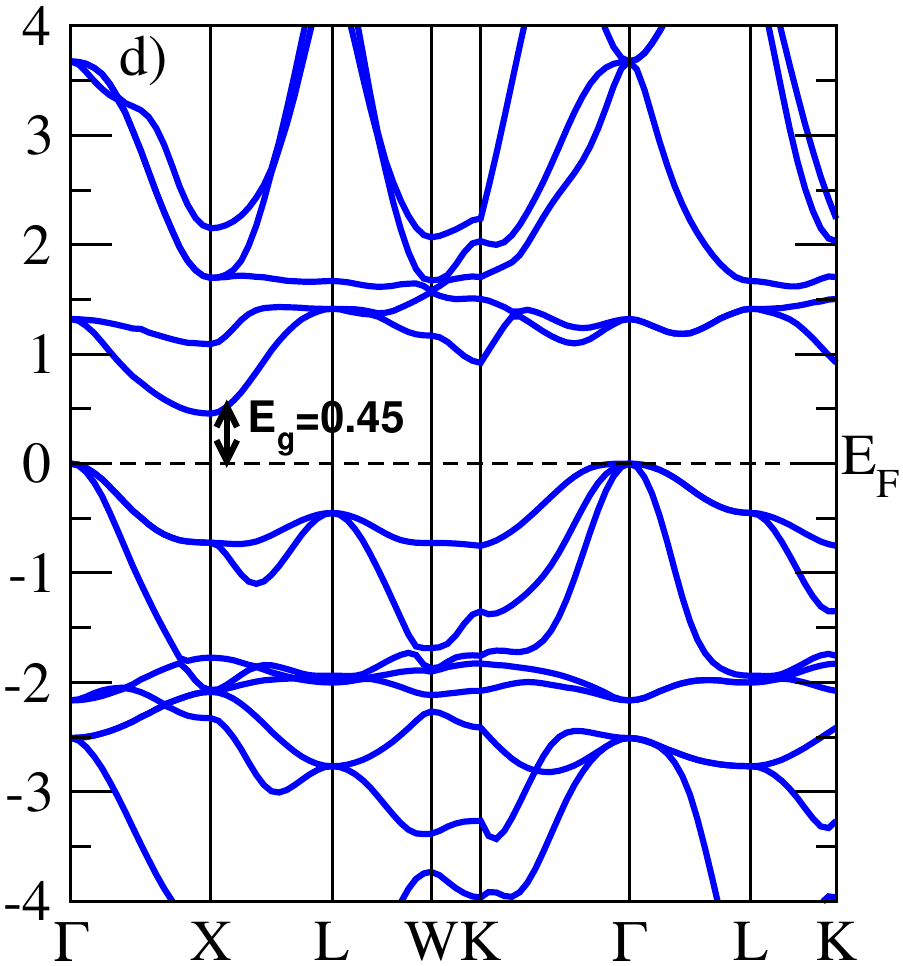}
	\caption{Electronic band structure of a) NiScP, b) NiScAs, c) NiScSb, and d) NiTiSn within GGA + mBJ.}
	\label{Fig4}
\end{figure}

It is interesting to note that with increase in the atomic radius of atoms at Z site, say, from P to Sb, the band gap reduces gradually which further leads to the decrease in the hybridization of Ni-$3d$ and Sc-$3d$ states. 	
An indirect band gap is observed in the band structures for hH alloys (see Figure \ref{Fig4}) with their valence band maximum (VBM) lying at $\Gamma$ and conduction band minimum (CBM) at X in the BZ.	
The VBM for the hH alloys are 3-fold degenerate comprising of heavy and light bands.	
From the observed band structure in Figure \ref{Fig4}, the scenario of heavy bands can enhance the Seebeck coefficient, whereas, the light band can facilitate the mobility of charge carriers \cite{kumarasinghe2019band,zhang2010zintl,fu2016enhancing}. 
Thus, the combination of heavy and light bands are preferable for increasing the TE performance. The band structure shown in Figure \ref{Fig4} (a), (b), and (c) dictates the effective mass to be more at $X-\Gamma$  in CBM than that of VBM at $\Gamma$ (i.e., the effective mass of electron at CBM is greater than that of the hole at VBM), which play an significant role in TE properties. 
As seen in NiTiSn (Figure \ref{Fig4} (d)), the VBM (at $\Gamma$) is flatter than the CBM (at $X$) indicating that the effective mass of holes at VBM is more than that of electrons on CBM.

\subsection{Transport Properties}
For an efficient TE material, a high value of $\alpha$ and $\sigma$  with a low  $\kappa$ is expected, as depicted in equation (\ref{eq1}). The dimensionless figure of merit $ZT$ can be optimized when these parameters are optimum. But these parameters are inter-related with themselves. Thus, to obtain high value of $ZT$ is in-sufficient just by tuning one or two parameters. To get insight into the TE properties of hH alloys, we calculate the Seebeck coefficient $\alpha$, electrical conductivity $\sigma$/$\tau$, thermal conductivity ($\kappa= \kappa_{e}+\kappa_{l}$), power factor (PF), and the $ZT$ by using constant relaxation time approximation and rigid band approximation. 

\begin{figure}[h!]
	\centering
	\includegraphics[width=\columnwidth,height=2in]{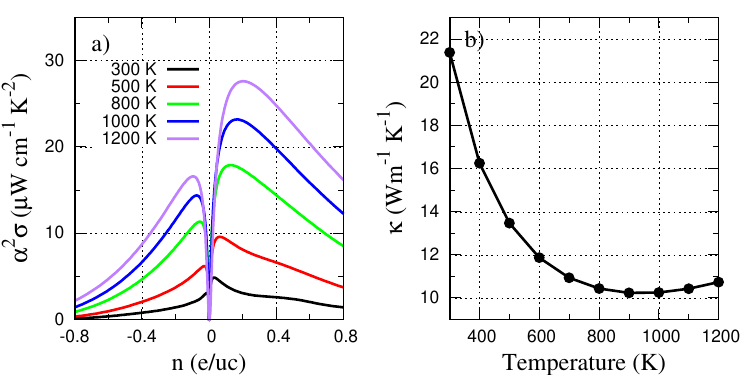}
	\caption{Power factor as a function of doping level (e/uc) for NiTiSn. The negative (positive) value represents the electron (hole) doping, and b) Total thermal conductivity as a function of temperature.}
	\label{Fig5}
\end{figure}

We first initiate our calculations for NiTiSn by validating the theoretical results, such as PF and thermal conductivity with the reported experimental measurements \cite{kim2007high}. From the comparison of the calculated and experimental electrical conductivity, we approximated the relaxation time $\tau$ = $\sim 2\times10^{-15}$ $s$. In the whole process, we use the constant relaxation time, even though it depends on doping level and temperature, obtained for NiTiSn to implement for all the iso-electronic systems. 

The PF of NiTiSn was reported to be $\sim 16$ $\mu$Wcm$^{-1}$K$^{-2}$ at 700 K, which upon electron doping (by 1\% of Sb atom to Sn site), PF rises to $\sim 30$ $\mu$W cm$^{-1}$K$^{-2}$. When temperature rises above 700 K, PF is found to decrease in both cases. Comparing these values we estimate that PF may range between $10-15$ $\mu$Wcm$^{-1}$K$^2$ at $0.04-0.06$ doping level of electron per unit cell in the same temperature range. 
In case of hole doping, PF lies within $17-23$ $\mu$W cm$^{-1}$K$^{-2}$ at the same temperature range when dopants is $0.1-0.2$ hole per unit cell. 
This indicates that hole doping is more appropriate than the electrons for PF. 
The calculated total thermal conductivity $21-10$ Wm$^{-1}K^{-1}$ (see Figure \ref{Fig5} (b)) was slightly higher than the earlier report (i.e., $7-10$ Wm$^{-1}K^{-1}$), which is mainly due to the electronic contribution found prominent at higher temperature. Our calculated results are comparable with the experimental measurements \cite{kim2007high}. 

\begin{figure}[h!]
	\centering
	\includegraphics[width=\columnwidth,height=4in]{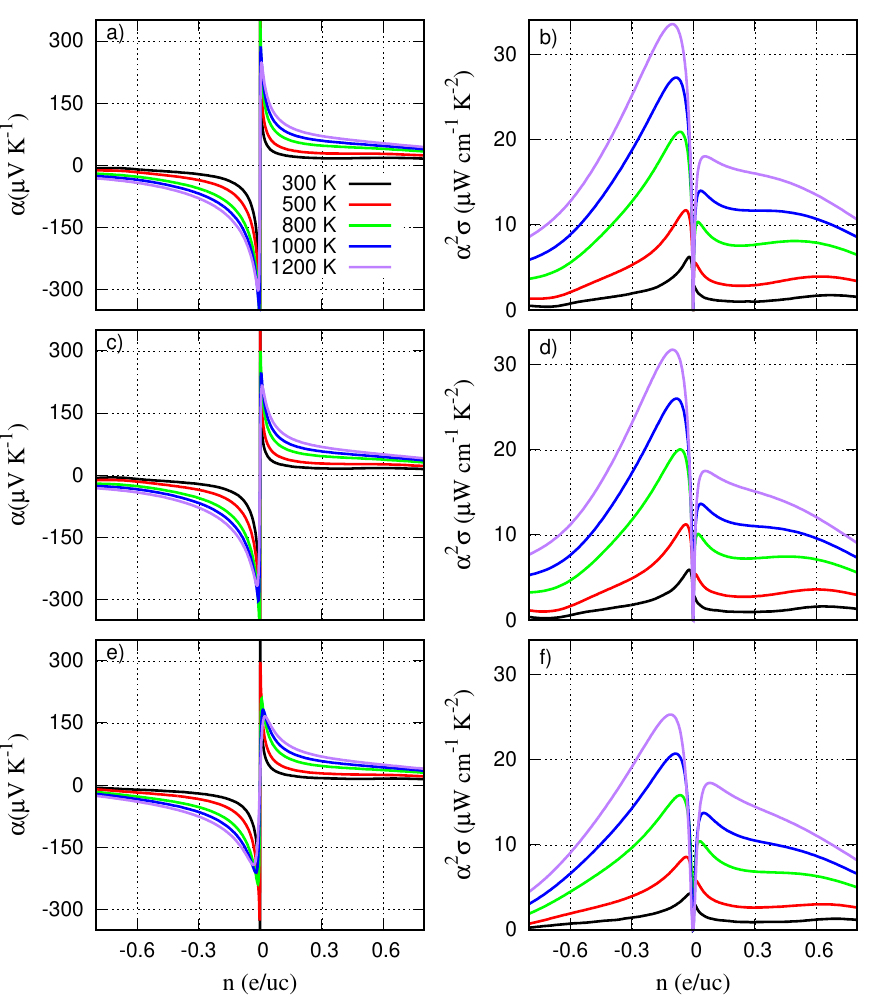}
	\caption{The Seebeck coefficient (a, c, e) and the Power factor (b, d, f) vs the doping level (in e/uc) at various temperature for NiScP, NiScAs, and NiScSb, respectively. The values in negative (positive) values on the horizontal axes represents the electron (hole) doping, respectively.}
	\label{Fig6}
\end{figure}

The Seebeck coefficient (a, c, e) and the PF (b, d, f) for different level of doping are shown in Figure \ref{Fig6} for NiScP, NiScAs, and NiScSb, respectively. Around $E_{\rm F}$ (i.e., at $\mu=0$), the Seebeck coefficient is large ($>\pm 150$ $\mu V/K$), which on doping to either side, falls-off significantly. This is evident from its inverse relation with the carrier concentration. 

The optimum values of the doping levels and corresponding TE parameters for 1200 K are listed in Table \ref{tab2}. 

\begin{figure}
	\includegraphics[width=\columnwidth]{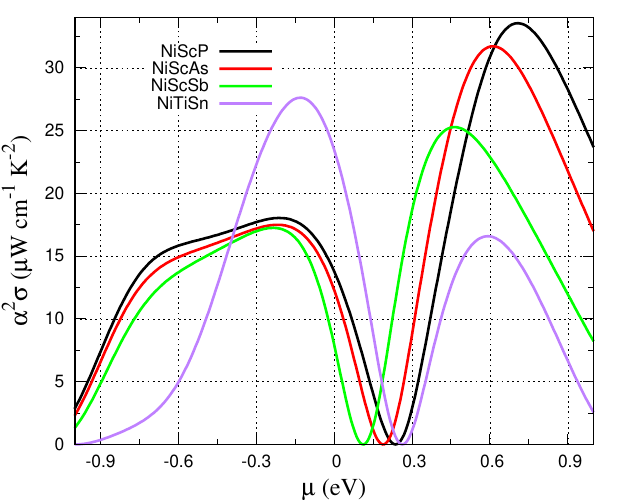}
	\caption{The power factor vs chemical potential ($\mu$)  at 1200 K temperature. The values of chemical potential in negative (positive) represents the hole (electron) doping.}
\label{TE_3}
\end{figure}

\begin{figure}[h!]
	\centering
	\includegraphics[width=3in,height=3in]{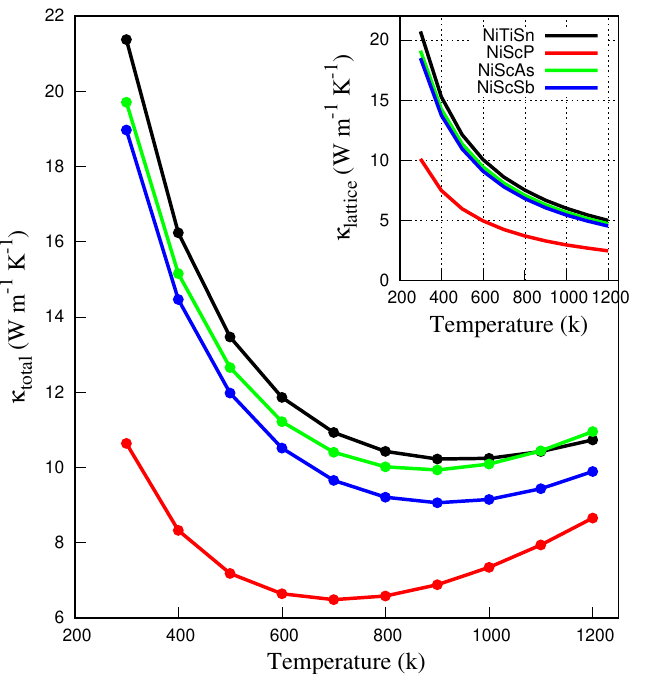}
	\caption{Total thermal conductivity ($\kappa$) as a function of temperature, inset dictates lattice contribution to thermal conductivity ($\kappa_l$) of NiScP, NiScAs, NiScSb, and NiTiSn.}
	\label{Fig7}
\end{figure}

PF is another parameter to check the reliability of TE materials. As observed in Figure \ref{Fig6}, the PF value for $p$ or $n-$ type is significant within the doping range of $\pm 0.3$. To be specific, at 1000 K the calculated values are approximately 15, 12 and 13 $\mu$Wcm$^{-1}$K$^{-2}$ for NiScP, NiScAs, and NiScSb within $0.02 - 0.04$ hole per unit cell reaching its maximum value at 1200 K. Similarly, for doping range $0.06 - 0.07$ electron per unit cell, PF rises to $\sim$ 27, 25, and 20 $\mu$Wcm$^{-1}$K$^{-2}$ at 1000 K, respectively. 
The sizable value of PF within the doping range $0.07 - 0.08$ electron per unit cell suggests that these material could be a good TE materials.

We further show the variation of PF with the chemical potential, $\mu$, in  Figure \ref{TE_3}. The peak values of PF noted in the chemical potential ranges between $0.4-0.7$ eV for NiScP, NiScAs and NiScSb. In contrast, the peak value of PF are around $-0.2$ eV for NiTiSn. 
From the above scenario, electron doping is found to be more suitable for NiScP, NiScAs, and NiScSb due to larger effective mass of electrons to get better TE performance. This indicates the presence of larger electron pockets resulting in the dense carriers which are confined on the CBM along $\Gamma-X$ (Fig. 4a-c). On the otherhand, in NiTiSn, hole doping is much more favorable due to the higher effective mass of holes resulting from the nearly flatter band in the VBM and CBM along $\Gamma-X$ (see band structure in Fig. 4d). 

\begin{table}
	\caption{\label{tabtwo}Calculated optimal doping levels and the corresponding Seebeck coefficient, electrical conductivity, power factor, and	$ZT$ of NiTZ (T= Sc, and Ti; Z= P, As, Sn, and Sb) in cubic symmetry $F\bar{4}3m$ at 1200 K. Negative ($-$) sign indicates the $n-$type characteristics.} 
	\centering
\begin{tabular}{cccccc}
	\hline\hline
	System & n&$\alpha$ & $\sigma$&$\alpha$$^2$$\sigma$& ZT\\
	&e/uc&$\mu$VK$^{-1}$&($\times$10$^3$ S cm$^{-1}$)&$\mu$Wcm$^{-1}$K$^{-2}$\\ 
	\hline \\
	NiTiSn  &0.20 &154 &1.15 &27.61&0.30 \\ \\
	NiScP  &-0.08 &-177&1.05 &33.16&0.46 \\ \\
	NiScAs &-0.08 &-168&1.12 &31.50&0.35\\ \\
	NiScSb &-0.07 &-163&0.90 &24.20&0.29 \\ 
	\hline\hline
\end{tabular}
\label{tab2}
\end{table}

\begin{figure}[h!]
	\centering
	\includegraphics[width=3in,height=3in]{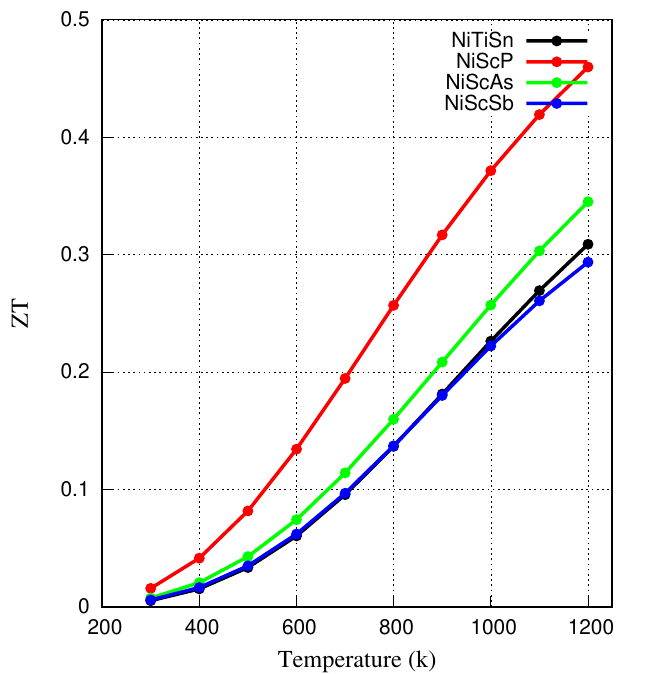}
	\caption{The $ZT$ as a function of temperature.}
	\label{Fig8}
\end{figure}

Figure \ref{Fig7} shows the calculated thermal conductivity as a function of temperature for NiScP, NiScAs, NiScSb, and NiTiSn, respectively. The total thermal conductivity consists of two components \textit{viz.} electronic ($\kappa_e$) and lattice ($\kappa_l$) parts. At low temperature (say 300 K), the lattice part was found dominant over the electronic part, and with rise in temperature (say upto $\sim 900$ K, except NiScP), the lattice thermal conductivity and the overall conductivity decreases uniformly. 

To note is, with an increase in temperature starting from 300 K, the carrier concentration increases resulting in higher electrical conductivity, and hence the overall thermal conductivity. Similar features was observed in the recent report of CoMnSb \cite{hori2020first}.
The calculated lattice conductivity are 10.6, 19, and 18.5 Wm$^{-1}K^{-1}$ at 300 K which reduces abruptly to 2.5, 4.7, and 4.5 Wm$^{-1}K^{-1}$ at 1200 K for NiScP, NiScAs, and NiScSb, respectively. 

The figure of merit $ZT$ for hH alloys is as shown in Figure \ref{Fig8}. With low value (say, 0.05) of $ZT$ at 300 K, it is found to rise linearly with the increase in temperature. At 1200  K, the calculated values are 0.30, 0.45, 0.35 and 0.29, respectively for NiTiSn, NiScP, NiScAs, and NiScSb alloys. On the otherhand, the values of total thermal conductivity is found minimum at $\sim 700$ K for NiScP and $\sim 900$ K for the remaining systems which starts increases afterwards due to the dominance of electronic part. The variation of $ZT$ with temperature shows linear nature. PF is dominant at higher temperature range due to the increase in carrier concentration along with the electrical conductivity.  
 The observed $ZT$ is low mainly due to a higher value of $\kappa$.  Even if our $ZT$ values are lower than the commercialized TE materials such as Bi$_2$Te$_3$ and PbTe but can be enhanced by means of doping to any of the three atomic sites. 
Though the $ZT$ value is low ($\sim0.05$) for the pristine systems compared to the widely used doped Bi-based alloys, say, Bi$_2$Se$_3$ ($\sim0.01-0.05$)\cite{mishra1997electronic}, Bi$_2$Se$_3$ at Bi$_2$Te$_3$ ($\sim0.7$)\cite{min2013surfactant}, and Bi-Sb alloys ($\sim0.4$)\cite{lenoir1996transport}, it can be enhanced by electron doping\cite{fang2020anisotropic}.

As observed from the calculations above, $ZT$ value can increase when PF is enhanced while minimizing the thermal conductivity. The possible route to tune this from DF is by proper tuning of the band gap with appropriate electron/hole doping as discussed.

\section{Conclusions}	
On the basis on density functional calculations, we investigate the half-Heuslers NiTiSn, NiScP, NiScAs, and NiScSb, respectively. Electronic properties reveal that these materials are semiconductor with an indirect band gap. The narrow-band gap marks them as suitable candidate for TE performance. The calculated power factor shows large value in both the electron and hole doping case. Electron doping is found more preferable than hole for NiScP, NiScAs, and NiScSb, while hole doping is preferable for NiTiSn. Based on the constant relaxation time approximation and rigid band approximation with sizable $ZT$, these compounds are predicted as a possible TE materials.

\begin{acknowledgments}
MPG acknowledges the Department of Science and Technology, India for awarding the India Science and Research Fellowship (ISRF-2019) with grant number DO/CCSTDS/201/2019, and Alexander von Humboldt Foundation, Germany for the equipment subsidy grants. Part of this work was performed at IIT-Roorkee, India during the ISRF-2019 program, and part of it with the computational resources provided by the Kathmandu University Supercomputer Center established with the equipment donated by CERN. 
MPG thanks H. C. Kandpal for fruitful discussions suggesting half-Heusler group to explore and also for all the logistic and technical supports at IIT-Roorkee.
\end{acknowledgments}

\section*{DATA AVAILABILITY}
The data that support the findings of this study are available with
the corresponding author and can be obtained upon reasonable
request.

\section*{Declaration of competing interest}
There is no conflict of interest.

\nocite{*}
\bibliography{references}

\begin{thebibliography}{64}%
\makeatletter
\providecommand \@ifxundefined [1]{%
 \@ifx{#1\undefined}
}%
\providecommand \@ifnum [1]{%
 \ifnum #1\expandafter \@firstoftwo
 \else \expandafter \@secondoftwo
 \fi
}%
\providecommand \@ifx [1]{%
 \ifx #1\expandafter \@firstoftwo
 \else \expandafter \@secondoftwo
 \fi
}%
\providecommand \natexlab [1]{#1}%
\providecommand \enquote  [1]{``#1''}%
\providecommand \bibnamefont  [1]{#1}%
\providecommand \bibfnamefont [1]{#1}%
\providecommand \citenamefont [1]{#1}%
\providecommand \href@noop [0]{\@secondoftwo}%
\providecommand \href [0]{\begingroup \@sanitize@url \@href}%
\providecommand \@href[1]{\@@startlink{#1}\@@href}%
\providecommand \@@href[1]{\endgroup#1\@@endlink}%
\providecommand \@sanitize@url [0]{\catcode `\\12\catcode `\$12\catcode
  `\&12\catcode `\#12\catcode `\^12\catcode `\_12\catcode `\%12\relax}%
\providecommand \@@startlink[1]{}%
\providecommand \@@endlink[0]{}%
\providecommand \url  [0]{\begingroup\@sanitize@url \@url }%
\providecommand \@url [1]{\endgroup\@href {#1}{\urlprefix }}%
\providecommand \urlprefix  [0]{URL }%
\providecommand \Eprint [0]{\href }%
\providecommand \doibase [0]{http://dx.doi.org/}%
\providecommand \selectlanguage [0]{\@gobble}%
\providecommand \bibinfo  [0]{\@secondoftwo}%
\providecommand \bibfield  [0]{\@secondoftwo}%
\providecommand \translation [1]{[#1]}%
\providecommand \BibitemOpen [0]{}%
\providecommand \bibitemStop [0]{}%
\providecommand \bibitemNoStop [0]{.\EOS\space}%
\providecommand \EOS [0]{\spacefactor3000\relax}%
\providecommand \BibitemShut  [1]{\csname bibitem#1\endcsname}%
\let\auto@bib@innerbib\@empty
\bibitem [{\citenamefont {Russ}\ \emph {et~al.}(2016)\citenamefont {Russ},
  \citenamefont {Glaudell}, \citenamefont {Urban}, \citenamefont {Chabinyc},\
  and\ \citenamefont {Segalman}}]{russ2016organic}%
  \BibitemOpen
  \bibfield  {author} {\bibinfo {author} {\bibfnamefont {B.}~\bibnamefont
  {Russ}}, \bibinfo {author} {\bibfnamefont {A.}~\bibnamefont {Glaudell}},
  \bibinfo {author} {\bibfnamefont {J.~J.}\ \bibnamefont {Urban}}, \bibinfo
  {author} {\bibfnamefont {M.~L.}\ \bibnamefont {Chabinyc}}, \ and\ \bibinfo
  {author} {\bibfnamefont {R.~A.}\ \bibnamefont {Segalman}},\ }\bibfield
  {title} {\enquote {\bibinfo {title} {Organic thermoelectric materials for
  energy harvesting and temperature control},}\ }\href@noop {} {\bibfield
  {journal} {\bibinfo  {journal} {Nature Reviews Materials}\ }\textbf {\bibinfo
  {volume} {1}},\ \bibinfo {pages} {1--14} (\bibinfo {year}
  {2016})}\BibitemShut {NoStop}%
\bibitem [{\citenamefont {Kanatzidis}(2010)}]{kanatzidis2010nanostructured}%
  \BibitemOpen
  \bibfield  {author} {\bibinfo {author} {\bibfnamefont {M.~G.}\ \bibnamefont
  {Kanatzidis}},\ }\bibfield  {title} {\enquote {\bibinfo {title}
  {Nanostructured thermoelectrics: the new paradigm?}}\ }\href@noop {}
  {\bibfield  {journal} {\bibinfo  {journal} {Chemistry of materials}\ }\textbf
  {\bibinfo {volume} {22}},\ \bibinfo {pages} {648--659} (\bibinfo {year}
  {2010})}\BibitemShut {NoStop}%
\bibitem [{\citenamefont {Snyder}\ and\ \citenamefont
  {Toberer}(2011)}]{snyder2011complex}%
  \BibitemOpen
  \bibfield  {author} {\bibinfo {author} {\bibfnamefont {G.~J.}\ \bibnamefont
  {Snyder}}\ and\ \bibinfo {author} {\bibfnamefont {E.~S.}\ \bibnamefont
  {Toberer}},\ }\bibfield  {title} {\enquote {\bibinfo {title} {Complex
  thermoelectric materials},}\ }in\ \href@noop {} {\emph {\bibinfo {booktitle}
  {materials for sustainable energy: a collection of peer-reviewed research and
  review articles from Nature Publishing Group}}}\ (\bibinfo  {publisher}
  {World Scientific},\ \bibinfo {year} {2011})\ pp.\ \bibinfo {pages}
  {101--110}\BibitemShut {NoStop}%
\bibitem [{\citenamefont {Lan}\ \emph {et~al.}(2010)\citenamefont {Lan},
  \citenamefont {Minnich}, \citenamefont {Chen},\ and\ \citenamefont
  {Ren}}]{lan2010enhancement}%
  \BibitemOpen
  \bibfield  {author} {\bibinfo {author} {\bibfnamefont {Y.}~\bibnamefont
  {Lan}}, \bibinfo {author} {\bibfnamefont {A.~J.}\ \bibnamefont {Minnich}},
  \bibinfo {author} {\bibfnamefont {G.}~\bibnamefont {Chen}}, \ and\ \bibinfo
  {author} {\bibfnamefont {Z.}~\bibnamefont {Ren}},\ }\bibfield  {title}
  {\enquote {\bibinfo {title} {Enhancement of thermoelectric figure-of-merit by
  a bulk nanostructuring approach},}\ }\href@noop {} {\bibfield  {journal}
  {\bibinfo  {journal} {Advanced Functional Materials}\ }\textbf {\bibinfo
  {volume} {20}},\ \bibinfo {pages} {357--376} (\bibinfo {year}
  {2010})}\BibitemShut {NoStop}%
\bibitem [{\citenamefont {Szczech}, \citenamefont {Higgins},\ and\
  \citenamefont {Jin}(2011)}]{szczech2011enhancement}%
  \BibitemOpen
  \bibfield  {author} {\bibinfo {author} {\bibfnamefont {J.~R.}\ \bibnamefont
  {Szczech}}, \bibinfo {author} {\bibfnamefont {J.~M.}\ \bibnamefont
  {Higgins}}, \ and\ \bibinfo {author} {\bibfnamefont {S.}~\bibnamefont
  {Jin}},\ }\bibfield  {title} {\enquote {\bibinfo {title} {Enhancement of the
  thermoelectric properties in nanoscale and nanostructured materials},}\
  }\href@noop {} {\bibfield  {journal} {\bibinfo  {journal} {Journal of
  Materials Chemistry}\ }\textbf {\bibinfo {volume} {21}},\ \bibinfo {pages}
  {4037--4055} (\bibinfo {year} {2011})}\BibitemShut {NoStop}%
\bibitem [{\citenamefont {Shankar}\ \emph {et~al.}(2017)\citenamefont
  {Shankar}, \citenamefont {Rai}, \citenamefont {Ghimire}, \citenamefont
  {Thapa} \emph {et~al.}}]{shankar2017electronic}%
  \BibitemOpen
  \bibfield  {author} {\bibinfo {author} {\bibfnamefont {A.}~\bibnamefont
  {Shankar}}, \bibinfo {author} {\bibfnamefont {D.}~\bibnamefont {Rai}},
  \bibinfo {author} {\bibfnamefont {M.}~\bibnamefont {Ghimire}}, \bibinfo
  {author} {\bibfnamefont {R.}~\bibnamefont {Thapa}},  \emph {et~al.},\
  }\bibfield  {title} {\enquote {\bibinfo {title} {Electronic structure and
  thermoelectricity of filled skutterudite euru 4 as 12: a dft calculation},}\
  }\href@noop {} {\bibfield  {journal} {\bibinfo  {journal} {Indian Journal of
  Physics}\ }\textbf {\bibinfo {volume} {91}},\ \bibinfo {pages} {17--23}
  (\bibinfo {year} {2017})}\BibitemShut {NoStop}%
\bibitem [{\citenamefont {Ghimire}, \citenamefont {Sandeep},\ and\
  \citenamefont {Thapa}(2010)}]{ghimire2010study}%
  \BibitemOpen
  \bibfield  {author} {\bibinfo {author} {\bibfnamefont {M.}~\bibnamefont
  {Ghimire}}, \bibinfo {author} {\bibnamefont {Sandeep}}, \ and\ \bibinfo
  {author} {\bibfnamefont {R.}~\bibnamefont {Thapa}},\ }\bibfield  {title}
  {\enquote {\bibinfo {title} {Study of the electronic properties of cro 2
  using density functional theory},}\ }\href@noop {} {\bibfield  {journal}
  {\bibinfo  {journal} {Modern Physics Letters B}\ }\textbf {\bibinfo {volume}
  {24}},\ \bibinfo {pages} {2187--2193} (\bibinfo {year} {2010})}\BibitemShut
  {NoStop}%
\bibitem [{\citenamefont {Ghimire}\ \emph {et~al.}(2015)\citenamefont
  {Ghimire}, \citenamefont {Thapa}, \citenamefont {Rai}, \citenamefont
  {Sandeep}, \citenamefont {Sinha},\ and\ \citenamefont
  {Hu}}]{ghimire2015half}%
  \BibitemOpen
  \bibfield  {author} {\bibinfo {author} {\bibfnamefont {M.~P.}\ \bibnamefont
  {Ghimire}}, \bibinfo {author} {\bibfnamefont {R.}~\bibnamefont {Thapa}},
  \bibinfo {author} {\bibfnamefont {D.}~\bibnamefont {Rai}}, \bibinfo {author}
  {\bibnamefont {Sandeep}}, \bibinfo {author} {\bibfnamefont {T.}~\bibnamefont
  {Sinha}}, \ and\ \bibinfo {author} {\bibfnamefont {X.}~\bibnamefont {Hu}},\
  }\bibfield  {title} {\enquote {\bibinfo {title} {Half metallic ferromagnetism
  in tri-layered perovskites sr4t3o10 (t= co, rh)},}\ }\href@noop {} {\bibfield
   {journal} {\bibinfo  {journal} {Journal of Applied Physics}\ }\textbf
  {\bibinfo {volume} {117}},\ \bibinfo {pages} {063903} (\bibinfo {year}
  {2015})}\BibitemShut {NoStop}%
\bibitem [{\citenamefont {Ghimire}\ and\ \citenamefont
  {Hu}(2016)}]{ghimire2016compensated}%
  \BibitemOpen
  \bibfield  {author} {\bibinfo {author} {\bibfnamefont {M.~P.}\ \bibnamefont
  {Ghimire}}\ and\ \bibinfo {author} {\bibfnamefont {X.}~\bibnamefont {Hu}},\
  }\bibfield  {title} {\enquote {\bibinfo {title} {Compensated half metallicity
  in osmium double perovskite driven by doping effects},}\ }\href@noop {}
  {\bibfield  {journal} {\bibinfo  {journal} {Materials Research Express}\
  }\textbf {\bibinfo {volume} {3}},\ \bibinfo {pages} {106107} (\bibinfo {year}
  {2016})}\BibitemShut {NoStop}%
\bibitem [{\citenamefont {Roy}, \citenamefont {Waghmare},\ and\ \citenamefont
  {Maiti}(2016)}]{roy2016environmentally}%
  \BibitemOpen
  \bibfield  {author} {\bibinfo {author} {\bibfnamefont {P.}~\bibnamefont
  {Roy}}, \bibinfo {author} {\bibfnamefont {V.}~\bibnamefont {Waghmare}}, \
  and\ \bibinfo {author} {\bibfnamefont {T.}~\bibnamefont {Maiti}},\ }\bibfield
   {title} {\enquote {\bibinfo {title} {Environmentally friendly ba x sr 2- x
  tifeo 6 double perovskite with enhanced thermopower for high temperature
  thermoelectric power generation},}\ }\href@noop {} {\bibfield  {journal}
  {\bibinfo  {journal} {RSC Advances}\ }\textbf {\bibinfo {volume} {6}},\
  \bibinfo {pages} {54636--54643} (\bibinfo {year} {2016})}\BibitemShut
  {NoStop}%
\bibitem [{\citenamefont {Choi}, \citenamefont {Yoo},\ and\ \citenamefont
  {Ohta}(2015)}]{choi2015polaron}%
  \BibitemOpen
  \bibfield  {author} {\bibinfo {author} {\bibfnamefont {W.~S.}\ \bibnamefont
  {Choi}}, \bibinfo {author} {\bibfnamefont {H.~K.}\ \bibnamefont {Yoo}}, \
  and\ \bibinfo {author} {\bibfnamefont {H.}~\bibnamefont {Ohta}},\ }\bibfield
  {title} {\enquote {\bibinfo {title} {Polaron transport and thermoelectric
  behavior in la-doped srtio3 thin films with elemental vacancies},}\
  }\href@noop {} {\bibfield  {journal} {\bibinfo  {journal} {Advanced
  Functional Materials}\ }\textbf {\bibinfo {volume} {25}},\ \bibinfo {pages}
  {799--804} (\bibinfo {year} {2015})}\BibitemShut {NoStop}%
\bibitem [{\citenamefont {Bhandari}\ \emph {et~al.}(2020)\citenamefont
  {Bhandari}, \citenamefont {Yadav}, \citenamefont {Belbase}, \citenamefont
  {Zeeshan}, \citenamefont {Sadhukhan}, \citenamefont {Rai}, \citenamefont
  {Thapa}, \citenamefont {Kaphle},\ and\ \citenamefont
  {Ghimire}}]{bhandari2020electronic}%
  \BibitemOpen
  \bibfield  {author} {\bibinfo {author} {\bibfnamefont {S.~R.}\ \bibnamefont
  {Bhandari}}, \bibinfo {author} {\bibfnamefont {D.}~\bibnamefont {Yadav}},
  \bibinfo {author} {\bibfnamefont {B.}~\bibnamefont {Belbase}}, \bibinfo
  {author} {\bibfnamefont {M.}~\bibnamefont {Zeeshan}}, \bibinfo {author}
  {\bibfnamefont {B.}~\bibnamefont {Sadhukhan}}, \bibinfo {author}
  {\bibfnamefont {D.}~\bibnamefont {Rai}}, \bibinfo {author} {\bibfnamefont
  {R.}~\bibnamefont {Thapa}}, \bibinfo {author} {\bibfnamefont
  {G.}~\bibnamefont {Kaphle}}, \ and\ \bibinfo {author} {\bibfnamefont {M.~P.}\
  \bibnamefont {Ghimire}},\ }\bibfield  {title} {\enquote {\bibinfo {title}
  {Electronic, magnetic, optical and thermoelectric properties of ca 2 cr 1- x
  ni x oso 6 double perovskites},}\ }\href@noop {} {\bibfield  {journal}
  {\bibinfo  {journal} {RSC Advances}\ }\textbf {\bibinfo {volume} {10}},\
  \bibinfo {pages} {16179--16186} (\bibinfo {year} {2020})}\BibitemShut
  {NoStop}%
\bibitem [{\citenamefont {Filippetti}\ \emph {et~al.}(2016)\citenamefont
  {Filippetti}, \citenamefont {Caddeo}, \citenamefont {Delugas},\ and\
  \citenamefont {Mattoni}}]{filippetti2016appealing}%
  \BibitemOpen
  \bibfield  {author} {\bibinfo {author} {\bibfnamefont {A.}~\bibnamefont
  {Filippetti}}, \bibinfo {author} {\bibfnamefont {C.}~\bibnamefont {Caddeo}},
  \bibinfo {author} {\bibfnamefont {P.}~\bibnamefont {Delugas}}, \ and\
  \bibinfo {author} {\bibfnamefont {A.}~\bibnamefont {Mattoni}},\ }\bibfield
  {title} {\enquote {\bibinfo {title} {Appealing perspectives of hybrid
  lead--iodide perovskites as thermoelectric materials},}\ }\href@noop {}
  {\bibfield  {journal} {\bibinfo  {journal} {The Journal of Physical Chemistry
  C}\ }\textbf {\bibinfo {volume} {120}},\ \bibinfo {pages} {28472--28479}
  (\bibinfo {year} {2016})}\BibitemShut {NoStop}%
\bibitem [{\citenamefont {Lee}\ \emph {et~al.}(2015)\citenamefont {Lee},
  \citenamefont {Hong}, \citenamefont {Stroppa}, \citenamefont {Whangbo},\ and\
  \citenamefont {Shim}}]{lee2015organic}%
  \BibitemOpen
  \bibfield  {author} {\bibinfo {author} {\bibfnamefont {C.}~\bibnamefont
  {Lee}}, \bibinfo {author} {\bibfnamefont {J.}~\bibnamefont {Hong}}, \bibinfo
  {author} {\bibfnamefont {A.}~\bibnamefont {Stroppa}}, \bibinfo {author}
  {\bibfnamefont {M.-H.}\ \bibnamefont {Whangbo}}, \ and\ \bibinfo {author}
  {\bibfnamefont {J.~H.}\ \bibnamefont {Shim}},\ }\bibfield  {title} {\enquote
  {\bibinfo {title} {Organic--inorganic hybrid perovskites abi 3 (a= ch 3 nh 3,
  nh 2 chnh 2; b= sn, pb) as potential thermoelectric materials: a density
  functional evaluation},}\ }\href@noop {} {\bibfield  {journal} {\bibinfo
  {journal} {RSC Advances}\ }\textbf {\bibinfo {volume} {5}},\ \bibinfo {pages}
  {78701--78707} (\bibinfo {year} {2015})}\BibitemShut {NoStop}%
\bibitem [{\citenamefont {Liu}\ \emph {et~al.}(2017)\citenamefont {Liu},
  \citenamefont {Li}, \citenamefont {Wang}, \citenamefont {Xu},\ and\
  \citenamefont {Hu}}]{liu2017extremely}%
  \BibitemOpen
  \bibfield  {author} {\bibinfo {author} {\bibfnamefont {Y.}~\bibnamefont
  {Liu}}, \bibinfo {author} {\bibfnamefont {X.}~\bibnamefont {Li}}, \bibinfo
  {author} {\bibfnamefont {J.}~\bibnamefont {Wang}}, \bibinfo {author}
  {\bibfnamefont {L.}~\bibnamefont {Xu}}, \ and\ \bibinfo {author}
  {\bibfnamefont {B.}~\bibnamefont {Hu}},\ }\bibfield  {title} {\enquote
  {\bibinfo {title} {An extremely high power factor in seebeck effects based on
  a new n-type copper-based organic/inorganic hybrid c 6 h 4 nh 2 cubr 2 i film
  with metal-like conductivity},}\ }\href@noop {} {\bibfield  {journal}
  {\bibinfo  {journal} {Journal of Materials Chemistry A}\ }\textbf {\bibinfo
  {volume} {5}},\ \bibinfo {pages} {13834--13841} (\bibinfo {year}
  {2017})}\BibitemShut {NoStop}%
\bibitem [{\citenamefont {Singh}\ \emph {et~al.}(2018)\citenamefont {Singh},
  \citenamefont {Wu}, \citenamefont {Yue}, \citenamefont {Romero},\ and\
  \citenamefont {Soluyanov}}]{PhysRevMaterials.2.114204}%
  \BibitemOpen
  \bibfield  {author} {\bibinfo {author} {\bibfnamefont {S.}~\bibnamefont
  {Singh}}, \bibinfo {author} {\bibfnamefont {Q.}~\bibnamefont {Wu}}, \bibinfo
  {author} {\bibfnamefont {C.}~\bibnamefont {Yue}}, \bibinfo {author}
  {\bibfnamefont {A.~H.}\ \bibnamefont {Romero}}, \ and\ \bibinfo {author}
  {\bibfnamefont {A.~A.}\ \bibnamefont {Soluyanov}},\ }\bibfield  {title}
  {\enquote {\bibinfo {title} {Topological phonons and thermoelectricity in
  triple-point metals},}\ }\href {\doibase 10.1103/PhysRevMaterials.2.114204}
  {\bibfield  {journal} {\bibinfo  {journal} {Phys. Rev. Materials}\ }\textbf
  {\bibinfo {volume} {2}},\ \bibinfo {pages} {114204} (\bibinfo {year}
  {2018})}\BibitemShut {NoStop}%
\bibitem [{\citenamefont {Rai}\ \emph {et~al.}(2017)\citenamefont {Rai},
  \citenamefont {Shankar}, \citenamefont {Sakhya}, \citenamefont {Sinha},
  \citenamefont {Grima-Gallardo}, \citenamefont {Cabrera}, \citenamefont
  {Khenata}, \citenamefont {Ghimire}, \citenamefont {Thapa} \emph
  {et~al.}}]{rai2017electronic}%
  \BibitemOpen
  \bibfield  {author} {\bibinfo {author} {\bibfnamefont {D.}~\bibnamefont
  {Rai}}, \bibinfo {author} {\bibfnamefont {A.}~\bibnamefont {Shankar}},
  \bibinfo {author} {\bibfnamefont {A.~P.}\ \bibnamefont {Sakhya}}, \bibinfo
  {author} {\bibfnamefont {T.}~\bibnamefont {Sinha}}, \bibinfo {author}
  {\bibfnamefont {P.}~\bibnamefont {Grima-Gallardo}}, \bibinfo {author}
  {\bibfnamefont {H.}~\bibnamefont {Cabrera}}, \bibinfo {author} {\bibfnamefont
  {R.}~\bibnamefont {Khenata}}, \bibinfo {author} {\bibfnamefont {M.~P.}\
  \bibnamefont {Ghimire}}, \bibinfo {author} {\bibfnamefont {R.}~\bibnamefont
  {Thapa}},  \emph {et~al.},\ }\bibfield  {title} {\enquote {\bibinfo {title}
  {Electronic, optical and thermoelectric properties of bulk and surface (001)
  cuinte2: A first principles study},}\ }\href@noop {} {\bibfield  {journal}
  {\bibinfo  {journal} {Journal of Alloys and Compounds}\ }\textbf {\bibinfo
  {volume} {699}},\ \bibinfo {pages} {1003--1011} (\bibinfo {year}
  {2017})}\BibitemShut {NoStop}%
\bibitem [{\citenamefont {Kim}, \citenamefont {Kimura},\ and\ \citenamefont
  {Mishima}(2007)}]{kim2007high}%
  \BibitemOpen
  \bibfield  {author} {\bibinfo {author} {\bibfnamefont {S.-W.}\ \bibnamefont
  {Kim}}, \bibinfo {author} {\bibfnamefont {Y.}~\bibnamefont {Kimura}}, \ and\
  \bibinfo {author} {\bibfnamefont {Y.}~\bibnamefont {Mishima}},\ }\bibfield
  {title} {\enquote {\bibinfo {title} {High temperature thermoelectric
  properties of tinisn-based half-heusler compounds},}\ }\href@noop {}
  {\bibfield  {journal} {\bibinfo  {journal} {Intermetallics}\ }\textbf
  {\bibinfo {volume} {15}},\ \bibinfo {pages} {349--356} (\bibinfo {year}
  {2007})}\BibitemShut {NoStop}%
\bibitem [{\citenamefont {Fu}\ \emph {et~al.}(2015)\citenamefont {Fu},
  \citenamefont {Bai}, \citenamefont {Liu}, \citenamefont {Tang}, \citenamefont
  {Chen}, \citenamefont {Zhao},\ and\ \citenamefont {Zhu}}]{fu2015realizing}%
  \BibitemOpen
  \bibfield  {author} {\bibinfo {author} {\bibfnamefont {C.}~\bibnamefont
  {Fu}}, \bibinfo {author} {\bibfnamefont {S.}~\bibnamefont {Bai}}, \bibinfo
  {author} {\bibfnamefont {Y.}~\bibnamefont {Liu}}, \bibinfo {author}
  {\bibfnamefont {Y.}~\bibnamefont {Tang}}, \bibinfo {author} {\bibfnamefont
  {L.}~\bibnamefont {Chen}}, \bibinfo {author} {\bibfnamefont {X.}~\bibnamefont
  {Zhao}}, \ and\ \bibinfo {author} {\bibfnamefont {T.}~\bibnamefont {Zhu}},\
  }\bibfield  {title} {\enquote {\bibinfo {title} {Realizing high figure of
  merit in heavy-band p-type half-heusler thermoelectric materials},}\
  }\href@noop {} {\bibfield  {journal} {\bibinfo  {journal} {Nature
  communications}\ }\textbf {\bibinfo {volume} {6}},\ \bibinfo {pages} {1--7}
  (\bibinfo {year} {2015})}\BibitemShut {NoStop}%
\bibitem [{\citenamefont {Zhu}\ \emph {et~al.}(2018)\citenamefont {Zhu},
  \citenamefont {He}, \citenamefont {Mao}, \citenamefont {Zhu}, \citenamefont
  {Li}, \citenamefont {Sun}, \citenamefont {Ren}, \citenamefont {Wang},
  \citenamefont {Liu}, \citenamefont {Tang} \emph {et~al.}}]{zhu2018discovery}%
  \BibitemOpen
  \bibfield  {author} {\bibinfo {author} {\bibfnamefont {H.}~\bibnamefont
  {Zhu}}, \bibinfo {author} {\bibfnamefont {R.}~\bibnamefont {He}}, \bibinfo
  {author} {\bibfnamefont {J.}~\bibnamefont {Mao}}, \bibinfo {author}
  {\bibfnamefont {Q.}~\bibnamefont {Zhu}}, \bibinfo {author} {\bibfnamefont
  {C.}~\bibnamefont {Li}}, \bibinfo {author} {\bibfnamefont {J.}~\bibnamefont
  {Sun}}, \bibinfo {author} {\bibfnamefont {W.}~\bibnamefont {Ren}}, \bibinfo
  {author} {\bibfnamefont {Y.}~\bibnamefont {Wang}}, \bibinfo {author}
  {\bibfnamefont {Z.}~\bibnamefont {Liu}}, \bibinfo {author} {\bibfnamefont
  {Z.}~\bibnamefont {Tang}},  \emph {et~al.},\ }\bibfield  {title} {\enquote
  {\bibinfo {title} {Discovery of zrcobi based half heuslers with high
  thermoelectric conversion efficiency},}\ }\href@noop {} {\bibfield  {journal}
  {\bibinfo  {journal} {Nature communications}\ }\textbf {\bibinfo {volume}
  {9}},\ \bibinfo {pages} {1--9} (\bibinfo {year} {2018})}\BibitemShut
  {NoStop}%
\bibitem [{\citenamefont {Zhu}\ \emph {et~al.}(2019)\citenamefont {Zhu},
  \citenamefont {Mao}, \citenamefont {Li}, \citenamefont {Sun}, \citenamefont
  {Wang}, \citenamefont {Zhu}, \citenamefont {Li}, \citenamefont {Song},
  \citenamefont {Zhou}, \citenamefont {Fu} \emph {et~al.}}]{zhu2019discovery}%
  \BibitemOpen
  \bibfield  {author} {\bibinfo {author} {\bibfnamefont {H.}~\bibnamefont
  {Zhu}}, \bibinfo {author} {\bibfnamefont {J.}~\bibnamefont {Mao}}, \bibinfo
  {author} {\bibfnamefont {Y.}~\bibnamefont {Li}}, \bibinfo {author}
  {\bibfnamefont {J.}~\bibnamefont {Sun}}, \bibinfo {author} {\bibfnamefont
  {Y.}~\bibnamefont {Wang}}, \bibinfo {author} {\bibfnamefont {Q.}~\bibnamefont
  {Zhu}}, \bibinfo {author} {\bibfnamefont {G.}~\bibnamefont {Li}}, \bibinfo
  {author} {\bibfnamefont {Q.}~\bibnamefont {Song}}, \bibinfo {author}
  {\bibfnamefont {J.}~\bibnamefont {Zhou}}, \bibinfo {author} {\bibfnamefont
  {Y.}~\bibnamefont {Fu}},  \emph {et~al.},\ }\bibfield  {title} {\enquote
  {\bibinfo {title} {Discovery of tafesb-based half-heuslers with high
  thermoelectric performance},}\ }\href@noop {} {\bibfield  {journal} {\bibinfo
   {journal} {Nature communications}\ }\textbf {\bibinfo {volume} {10}},\
  \bibinfo {pages} {1--8} (\bibinfo {year} {2019})}\BibitemShut {NoStop}%
\bibitem [{\citenamefont {Graf}\ \emph {et~al.}(2010)\citenamefont {Graf},
  \citenamefont {Klaer}, \citenamefont {Barth}, \citenamefont {Balke},
  \citenamefont {Elmers},\ and\ \citenamefont {Felser}}]{graf2010phase}%
  \BibitemOpen
  \bibfield  {author} {\bibinfo {author} {\bibfnamefont {T.}~\bibnamefont
  {Graf}}, \bibinfo {author} {\bibfnamefont {P.}~\bibnamefont {Klaer}},
  \bibinfo {author} {\bibfnamefont {J.}~\bibnamefont {Barth}}, \bibinfo
  {author} {\bibfnamefont {B.}~\bibnamefont {Balke}}, \bibinfo {author}
  {\bibfnamefont {H.-J.}\ \bibnamefont {Elmers}}, \ and\ \bibinfo {author}
  {\bibfnamefont {C.}~\bibnamefont {Felser}},\ }\bibfield  {title} {\enquote
  {\bibinfo {title} {Phase separation in the quaternary heusler compound coti
  (1- x) mnxsb a reduction in the thermal conductivity for thermoelectric
  applications},}\ }\href@noop {} {\bibfield  {journal} {\bibinfo  {journal}
  {Scripta Materialia}\ }\textbf {\bibinfo {volume} {63}},\ \bibinfo {pages}
  {1216--1219} (\bibinfo {year} {2010})}\BibitemShut {NoStop}%
\bibitem [{\citenamefont {Lee}, \citenamefont {Poudeu},\ and\ \citenamefont
  {Mahanti}(2011)}]{PhysRevB.83.085204}%
  \BibitemOpen
  \bibfield  {author} {\bibinfo {author} {\bibfnamefont {M.-S.}\ \bibnamefont
  {Lee}}, \bibinfo {author} {\bibfnamefont {F.~P.}\ \bibnamefont {Poudeu}}, \
  and\ \bibinfo {author} {\bibfnamefont {S.~D.}\ \bibnamefont {Mahanti}},\
  }\bibfield  {title} {\enquote {\bibinfo {title} {Electronic structure and
  thermoelectric properties of sb-based semiconducting half-heusler
  compounds},}\ }\href {\doibase 10.1103/PhysRevB.83.085204} {\bibfield
  {journal} {\bibinfo  {journal} {Phys. Rev. B}\ }\textbf {\bibinfo {volume}
  {83}},\ \bibinfo {pages} {085204} (\bibinfo {year} {2011})}\BibitemShut
  {NoStop}%
\bibitem [{\citenamefont {Zeeshan}\ \emph {et~al.}(2018)\citenamefont
  {Zeeshan}, \citenamefont {Nautiyal}, \citenamefont {van~den Brink},\ and\
  \citenamefont {Kandpal}}]{zeeshan2018fetasb}%
  \BibitemOpen
  \bibfield  {author} {\bibinfo {author} {\bibfnamefont {M.}~\bibnamefont
  {Zeeshan}}, \bibinfo {author} {\bibfnamefont {T.}~\bibnamefont {Nautiyal}},
  \bibinfo {author} {\bibfnamefont {J.}~\bibnamefont {van~den Brink}}, \ and\
  \bibinfo {author} {\bibfnamefont {H.~C.}\ \bibnamefont {Kandpal}},\
  }\bibfield  {title} {\enquote {\bibinfo {title} {Fetasb and femntisb as
  promising thermoelectric materials: An ab initio approach},}\ }\href@noop {}
  {\bibfield  {journal} {\bibinfo  {journal} {Physical Review Materials}\
  }\textbf {\bibinfo {volume} {2}},\ \bibinfo {pages} {065407} (\bibinfo {year}
  {2018})}\BibitemShut {NoStop}%
\bibitem [{\citenamefont {Singh}\ \emph
  {et~al.}(2019{\natexlab{a}})\citenamefont {Singh}, \citenamefont {Zeeshan},
  \citenamefont {Singh}, \citenamefont {van~den Brink},\ and\ \citenamefont
  {Kandpal}}]{singh2019first}%
  \BibitemOpen
  \bibfield  {author} {\bibinfo {author} {\bibfnamefont {S.}~\bibnamefont
  {Singh}}, \bibinfo {author} {\bibfnamefont {M.}~\bibnamefont {Zeeshan}},
  \bibinfo {author} {\bibfnamefont {U.}~\bibnamefont {Singh}}, \bibinfo
  {author} {\bibfnamefont {J.}~\bibnamefont {van~den Brink}}, \ and\ \bibinfo
  {author} {\bibfnamefont {H.~C.}\ \bibnamefont {Kandpal}},\ }\bibfield
  {title} {\enquote {\bibinfo {title} {First-principles investigations of
  orthorhombic-cubic phase transition and its effect on thermoelectric
  properties in cobalt-based ternary alloys},}\ }\href@noop {} {\bibfield
  {journal} {\bibinfo  {journal} {Journal of Physics: Condensed Matter}\
  }\textbf {\bibinfo {volume} {32}},\ \bibinfo {pages} {055505} (\bibinfo
  {year} {2019}{\natexlab{a}})}\BibitemShut {NoStop}%
\bibitem [{\citenamefont {Zeeshan}\ \emph {et~al.}(2017)\citenamefont
  {Zeeshan}, \citenamefont {Singh}, \citenamefont {van~den Brink},\ and\
  \citenamefont {Kandpal}}]{PhysRevMaterials.1.075407}%
  \BibitemOpen
  \bibfield  {author} {\bibinfo {author} {\bibfnamefont {M.}~\bibnamefont
  {Zeeshan}}, \bibinfo {author} {\bibfnamefont {H.~K.}\ \bibnamefont {Singh}},
  \bibinfo {author} {\bibfnamefont {J.}~\bibnamefont {van~den Brink}}, \ and\
  \bibinfo {author} {\bibfnamefont {H.~C.}\ \bibnamefont {Kandpal}},\
  }\bibfield  {title} {\enquote {\bibinfo {title} {Ab initio design of new
  cobalt-based half-heusler materials for thermoelectric applications},}\
  }\href {\doibase 10.1103/PhysRevMaterials.1.075407} {\bibfield  {journal}
  {\bibinfo  {journal} {Phys. Rev. Materials}\ }\textbf {\bibinfo {volume}
  {1}},\ \bibinfo {pages} {075407} (\bibinfo {year} {2017})}\BibitemShut
  {NoStop}%
\bibitem [{\citenamefont {Zeeshan}, \citenamefont {van~den Brink},\ and\
  \citenamefont {Kandpal}(2017)}]{PhysRevMaterials.1.074401}%
  \BibitemOpen
  \bibfield  {author} {\bibinfo {author} {\bibfnamefont {M.}~\bibnamefont
  {Zeeshan}}, \bibinfo {author} {\bibfnamefont {J.}~\bibnamefont {van~den
  Brink}}, \ and\ \bibinfo {author} {\bibfnamefont {H.~C.}\ \bibnamefont
  {Kandpal}},\ }\bibfield  {title} {\enquote {\bibinfo {title} {Hole-doped
  cobalt-based heusler phases as prospective high-performance high-temperature
  thermoelectrics},}\ }\href {\doibase 10.1103/PhysRevMaterials.1.074401}
  {\bibfield  {journal} {\bibinfo  {journal} {Phys. Rev. Materials}\ }\textbf
  {\bibinfo {volume} {1}},\ \bibinfo {pages} {074401} (\bibinfo {year}
  {2017})}\BibitemShut {NoStop}%
\bibitem [{\citenamefont {Singh}\ \emph
  {et~al.}(2019{\natexlab{b}})\citenamefont {Singh}, \citenamefont {Zeeshan},
  \citenamefont {Brink},\ and\ \citenamefont {Kandpal}}]{singh2019textit}%
  \BibitemOpen
  \bibfield  {author} {\bibinfo {author} {\bibfnamefont {S.}~\bibnamefont
  {Singh}}, \bibinfo {author} {\bibfnamefont {M.}~\bibnamefont {Zeeshan}},
  \bibinfo {author} {\bibfnamefont {J.~v.~d.}\ \bibnamefont {Brink}}, \ and\
  \bibinfo {author} {\bibfnamefont {H.~C.}\ \bibnamefont {Kandpal}},\
  }\bibfield  {title} {\enquote {\bibinfo {title} {Ab initio study of bi-based
  half heusler alloys as potential thermoelectric prospects},}\ }\href@noop {}
  {\bibfield  {journal} {\bibinfo  {journal} {arXiv:1904.02488}\ } (\bibinfo
  {year} {2019}{\natexlab{b}})}\BibitemShut {NoStop}%
\bibitem [{\citenamefont {Rai}\ \emph {et~al.}(2015)\citenamefont {Rai},
  \citenamefont {Shankar}, \citenamefont {Ghimire}, \citenamefont {Khenata},
  \citenamefont {Thapa} \emph {et~al.}}]{rai2015study}%
  \BibitemOpen
  \bibfield  {author} {\bibinfo {author} {\bibfnamefont {D.}~\bibnamefont
  {Rai}}, \bibinfo {author} {\bibfnamefont {A.}~\bibnamefont {Shankar}},
  \bibinfo {author} {\bibfnamefont {M.}~\bibnamefont {Ghimire}}, \bibinfo
  {author} {\bibfnamefont {R.}~\bibnamefont {Khenata}}, \bibinfo {author}
  {\bibfnamefont {R.}~\bibnamefont {Thapa}},  \emph {et~al.},\ }\bibfield
  {title} {\enquote {\bibinfo {title} {Study of the enhanced electronic and
  thermoelectric (te) properties of zrxhf1- x- ytaynisn: a first principles
  study},}\ }\href@noop {} {\bibfield  {journal} {\bibinfo  {journal} {RSC
  Advances}\ }\textbf {\bibinfo {volume} {5}},\ \bibinfo {pages} {95353--95359}
  (\bibinfo {year} {2015})}\BibitemShut {NoStop}%
\bibitem [{\citenamefont {Sakurada}\ and\ \citenamefont
  {Shutoh}(2005)}]{sakurada2005effect}%
  \BibitemOpen
  \bibfield  {author} {\bibinfo {author} {\bibfnamefont {S.}~\bibnamefont
  {Sakurada}}\ and\ \bibinfo {author} {\bibfnamefont {N.}~\bibnamefont
  {Shutoh}},\ }\bibfield  {title} {\enquote {\bibinfo {title} {Effect of ti
  substitution on the thermoelectric properties of (zr, hf) nisn half-heusler
  compounds},}\ }\href@noop {} {\bibfield  {journal} {\bibinfo  {journal}
  {Applied Physics Letters}\ }\textbf {\bibinfo {volume} {86}},\ \bibinfo
  {pages} {082105} (\bibinfo {year} {2005})}\BibitemShut {NoStop}%
\bibitem [{\citenamefont {Roy}\ \emph {et~al.}(2012)\citenamefont {Roy},
  \citenamefont {Bennett}, \citenamefont {Rabe},\ and\ \citenamefont
  {Vanderbilt}}]{PhysRevLett.109.037602}%
  \BibitemOpen
  \bibfield  {author} {\bibinfo {author} {\bibfnamefont {A.}~\bibnamefont
  {Roy}}, \bibinfo {author} {\bibfnamefont {J.~W.}\ \bibnamefont {Bennett}},
  \bibinfo {author} {\bibfnamefont {K.~M.}\ \bibnamefont {Rabe}}, \ and\
  \bibinfo {author} {\bibfnamefont {D.}~\bibnamefont {Vanderbilt}},\ }\bibfield
   {title} {\enquote {\bibinfo {title} {Half-heusler semiconductors as
  piezoelectrics},}\ }\href {\doibase 10.1103/PhysRevLett.109.037602}
  {\bibfield  {journal} {\bibinfo  {journal} {Phys. Rev. Lett.}\ }\textbf
  {\bibinfo {volume} {109}},\ \bibinfo {pages} {037602} (\bibinfo {year}
  {2012})}\BibitemShut {NoStop}%
\bibitem [{\citenamefont {Liu}\ \emph {et~al.}(2004)\citenamefont {Liu},
  \citenamefont {Hu}, \citenamefont {Liu}, \citenamefont {Cui}, \citenamefont
  {Zhang}, \citenamefont {Chen}, \citenamefont {Wu},\ and\ \citenamefont
  {Xiao}}]{PhysRevB.69.134415}%
  \BibitemOpen
  \bibfield  {author} {\bibinfo {author} {\bibfnamefont {Z.~H.}\ \bibnamefont
  {Liu}}, \bibinfo {author} {\bibfnamefont {H.~N.}\ \bibnamefont {Hu}},
  \bibinfo {author} {\bibfnamefont {G.~D.}\ \bibnamefont {Liu}}, \bibinfo
  {author} {\bibfnamefont {Y.~T.}\ \bibnamefont {Cui}}, \bibinfo {author}
  {\bibfnamefont {M.}~\bibnamefont {Zhang}}, \bibinfo {author} {\bibfnamefont
  {J.~L.}\ \bibnamefont {Chen}}, \bibinfo {author} {\bibfnamefont {G.~H.}\
  \bibnamefont {Wu}}, \ and\ \bibinfo {author} {\bibfnamefont {G.}~\bibnamefont
  {Xiao}},\ }\bibfield  {title} {\enquote {\bibinfo {title} {Electronic
  structure and ferromagnetism in the martensitic-transformation material
  ni2fega},}\ }\href {\doibase 10.1103/PhysRevB.69.134415} {\bibfield
  {journal} {\bibinfo  {journal} {Phys. Rev. B}\ }\textbf {\bibinfo {volume}
  {69}},\ \bibinfo {pages} {134415} (\bibinfo {year} {2004})}\BibitemShut
  {NoStop}%
\bibitem [{\citenamefont {Feng}\ \emph {et~al.}(2010)\citenamefont {Feng},
  \citenamefont {Xiao}, \citenamefont {Zhang},\ and\ \citenamefont
  {Yao}}]{PhysRevB.82.235121}%
  \BibitemOpen
  \bibfield  {author} {\bibinfo {author} {\bibfnamefont {W.}~\bibnamefont
  {Feng}}, \bibinfo {author} {\bibfnamefont {D.}~\bibnamefont {Xiao}}, \bibinfo
  {author} {\bibfnamefont {Y.}~\bibnamefont {Zhang}}, \ and\ \bibinfo {author}
  {\bibfnamefont {Y.}~\bibnamefont {Yao}},\ }\bibfield  {title} {\enquote
  {\bibinfo {title} {Half-heusler topological insulators: A first-principles
  study with the tran-blaha modified becke-johnson density functional},}\
  }\href {\doibase 10.1103/PhysRevB.82.235121} {\bibfield  {journal} {\bibinfo
  {journal} {Phys. Rev. B}\ }\textbf {\bibinfo {volume} {82}},\ \bibinfo
  {pages} {235121} (\bibinfo {year} {2010})}\BibitemShut {NoStop}%
\bibitem [{\citenamefont {de~Groot}\ \emph {et~al.}(1983)\citenamefont
  {de~Groot}, \citenamefont {Mueller}, \citenamefont {Engen},\ and\
  \citenamefont {Buschow}}]{PhysRevLett.50.2024}%
  \BibitemOpen
  \bibfield  {author} {\bibinfo {author} {\bibfnamefont {R.~A.}\ \bibnamefont
  {de~Groot}}, \bibinfo {author} {\bibfnamefont {F.~M.}\ \bibnamefont
  {Mueller}}, \bibinfo {author} {\bibfnamefont {P.~G.~v.}\ \bibnamefont
  {Engen}}, \ and\ \bibinfo {author} {\bibfnamefont {K.~H.~J.}\ \bibnamefont
  {Buschow}},\ }\bibfield  {title} {\enquote {\bibinfo {title} {New class of
  materials: Half-metallic ferromagnets},}\ }\href {\doibase
  10.1103/PhysRevLett.50.2024} {\bibfield  {journal} {\bibinfo  {journal}
  {Phys. Rev. Lett.}\ }\textbf {\bibinfo {volume} {50}},\ \bibinfo {pages}
  {2024--2027} (\bibinfo {year} {1983})}\BibitemShut {NoStop}%
\bibitem [{\citenamefont {Ghimire}\ \emph {et~al.}(2011)\citenamefont
  {Ghimire}, \citenamefont {Sandeep}, \citenamefont {Sinha},\ and\
  \citenamefont {Thapa}}]{ghimire2011first}%
  \BibitemOpen
  \bibfield  {author} {\bibinfo {author} {\bibfnamefont {M.}~\bibnamefont
  {Ghimire}}, \bibinfo {author} {\bibnamefont {Sandeep}}, \bibinfo {author}
  {\bibfnamefont {T.}~\bibnamefont {Sinha}}, \ and\ \bibinfo {author}
  {\bibfnamefont {R.}~\bibnamefont {Thapa}},\ }\bibfield  {title} {\enquote
  {\bibinfo {title} {First principles study of the electronic and magnetic
  properties of semi-heusler alloys nixsb (x= ti, v, cr and mn)},}\ }\href@noop
  {} {\bibfield  {journal} {\bibinfo  {journal} {Journal of alloys and
  compounds}\ }\textbf {\bibinfo {volume} {509}},\ \bibinfo {pages}
  {9742--9752} (\bibinfo {year} {2011})}\BibitemShut {NoStop}%
\bibitem [{\citenamefont {Sandeep}\ \emph {et~al.}(2012)\citenamefont
  {Sandeep}, \citenamefont {Ghimire}, \citenamefont {Deka}, \citenamefont
  {Rai}, \citenamefont {Shankar},\ and\ \citenamefont
  {Thapa}}]{ghimire2012magnetic}%
  \BibitemOpen
  \bibfield  {author} {\bibinfo {author} {\bibnamefont {Sandeep}}, \bibinfo
  {author} {\bibfnamefont {M.}~\bibnamefont {Ghimire}}, \bibinfo {author}
  {\bibfnamefont {D.}~\bibnamefont {Deka}}, \bibinfo {author} {\bibfnamefont
  {D.}~\bibnamefont {Rai}}, \bibinfo {author} {\bibfnamefont {A.}~\bibnamefont
  {Shankar}}, \ and\ \bibinfo {author} {\bibfnamefont {R.}~\bibnamefont
  {Thapa}},\ }\bibfield  {title} {\enquote {\bibinfo {title} {Magnetic and
  electronic properties of half-metallic nitbsb: a first principles study},}\
  }\href@noop {} {\bibfield  {journal} {\bibinfo  {journal} {Indian Journal of
  Physics}\ }\textbf {\bibinfo {volume} {86}},\ \bibinfo {pages} {301--305}
  (\bibinfo {year} {2012})}\BibitemShut {NoStop}%
\bibitem [{\citenamefont {Nakajima}\ \emph {et~al.}(2015)\citenamefont
  {Nakajima}, \citenamefont {Hu}, \citenamefont {Kirshenbaum}, \citenamefont
  {Hughes}, \citenamefont {Syers}, \citenamefont {Wang}, \citenamefont {Wang},
  \citenamefont {Wang}, \citenamefont {Saha}, \citenamefont {Pratt} \emph
  {et~al.}}]{nakajima2015topological}%
  \BibitemOpen
  \bibfield  {author} {\bibinfo {author} {\bibfnamefont {Y.}~\bibnamefont
  {Nakajima}}, \bibinfo {author} {\bibfnamefont {R.}~\bibnamefont {Hu}},
  \bibinfo {author} {\bibfnamefont {K.}~\bibnamefont {Kirshenbaum}}, \bibinfo
  {author} {\bibfnamefont {A.}~\bibnamefont {Hughes}}, \bibinfo {author}
  {\bibfnamefont {P.}~\bibnamefont {Syers}}, \bibinfo {author} {\bibfnamefont
  {X.}~\bibnamefont {Wang}}, \bibinfo {author} {\bibfnamefont {K.}~\bibnamefont
  {Wang}}, \bibinfo {author} {\bibfnamefont {R.}~\bibnamefont {Wang}}, \bibinfo
  {author} {\bibfnamefont {S.~R.}\ \bibnamefont {Saha}}, \bibinfo {author}
  {\bibfnamefont {D.}~\bibnamefont {Pratt}},  \emph {et~al.},\ }\bibfield
  {title} {\enquote {\bibinfo {title} {Topological rpdbi half-heusler
  semimetals: A new family of noncentrosymmetric magnetic superconductors},}\
  }\href@noop {} {\bibfield  {journal} {\bibinfo  {journal} {Science advances}\
  }\textbf {\bibinfo {volume} {1}},\ \bibinfo {pages} {e1500242} (\bibinfo
  {year} {2015})}\BibitemShut {NoStop}%
\bibitem [{\citenamefont {Zhang}\ \emph {et~al.}(2020)\citenamefont {Zhang},
  \citenamefont {Chen}, \citenamefont {Li}, \citenamefont {Zhang},
  \citenamefont {Hou}, \citenamefont {Wen}, \citenamefont {Zhang},
  \citenamefont {Wang},\ and\ \citenamefont {Zhang}}]{zhang2020topological}%
  \BibitemOpen
  \bibfield  {author} {\bibinfo {author} {\bibfnamefont {J.}~\bibnamefont
  {Zhang}}, \bibinfo {author} {\bibfnamefont {J.}~\bibnamefont {Chen}},
  \bibinfo {author} {\bibfnamefont {P.}~\bibnamefont {Li}}, \bibinfo {author}
  {\bibfnamefont {C.}~\bibnamefont {Zhang}}, \bibinfo {author} {\bibfnamefont
  {Z.}~\bibnamefont {Hou}}, \bibinfo {author} {\bibfnamefont {Y.}~\bibnamefont
  {Wen}}, \bibinfo {author} {\bibfnamefont {Q.}~\bibnamefont {Zhang}}, \bibinfo
  {author} {\bibfnamefont {W.}~\bibnamefont {Wang}}, \ and\ \bibinfo {author}
  {\bibfnamefont {X.}~\bibnamefont {Zhang}},\ }\bibfield  {title} {\enquote
  {\bibinfo {title} {Topological electronic state and anisotropic fermi surface
  in half-heusler gdptbi},}\ }\href@noop {} {\bibfield  {journal} {\bibinfo
  {journal} {Journal of Physics: Condensed Matter}\ }\textbf {\bibinfo {volume}
  {32}},\ \bibinfo {pages} {355707} (\bibinfo {year} {2020})}\BibitemShut
  {NoStop}%
\bibitem [{\citenamefont {Zaitsev}\ \emph {et~al.}(2006)\citenamefont
  {Zaitsev}, \citenamefont {Fedorov}, \citenamefont {Eremin}, \citenamefont
  {Gurieva},\ and\ \citenamefont {Rowe}}]{zaitsev2006thermoelectrics}%
  \BibitemOpen
  \bibfield  {author} {\bibinfo {author} {\bibfnamefont {V.}~\bibnamefont
  {Zaitsev}}, \bibinfo {author} {\bibfnamefont {M.}~\bibnamefont {Fedorov}},
  \bibinfo {author} {\bibfnamefont {I.}~\bibnamefont {Eremin}}, \bibinfo
  {author} {\bibfnamefont {E.}~\bibnamefont {Gurieva}}, \ and\ \bibinfo
  {author} {\bibfnamefont {D.}~\bibnamefont {Rowe}},\ }\bibfield  {title}
  {\enquote {\bibinfo {title} {Thermoelectrics handbook: macro to nano},}\
  }\href@noop {} {\bibfield  {journal} {\bibinfo  {journal} {CRC Press, Taylor
  \& Francis, Boca Raton}\ } (\bibinfo {year} {2006})}\BibitemShut {NoStop}%
\bibitem [{\citenamefont {Riffat}\ and\ \citenamefont
  {Ma}(2003)}]{riffat2003thermoelectrics}%
  \BibitemOpen
  \bibfield  {author} {\bibinfo {author} {\bibfnamefont {S.~B.}\ \bibnamefont
  {Riffat}}\ and\ \bibinfo {author} {\bibfnamefont {X.}~\bibnamefont {Ma}},\
  }\bibfield  {title} {\enquote {\bibinfo {title} {Thermoelectrics: a review of
  present and potential applications},}\ }\href@noop {} {\bibfield  {journal}
  {\bibinfo  {journal} {Applied thermal engineering}\ }\textbf {\bibinfo
  {volume} {23}},\ \bibinfo {pages} {913--935} (\bibinfo {year}
  {2003})}\BibitemShut {NoStop}%
\bibitem [{\citenamefont {Pei}\ \emph {et~al.}(2011)\citenamefont {Pei},
  \citenamefont {Shi}, \citenamefont {LaLonde}, \citenamefont {Wang},
  \citenamefont {Chen},\ and\ \citenamefont {Snyder}}]{pei2011convergence}%
  \BibitemOpen
  \bibfield  {author} {\bibinfo {author} {\bibfnamefont {Y.}~\bibnamefont
  {Pei}}, \bibinfo {author} {\bibfnamefont {X.}~\bibnamefont {Shi}}, \bibinfo
  {author} {\bibfnamefont {A.}~\bibnamefont {LaLonde}}, \bibinfo {author}
  {\bibfnamefont {H.}~\bibnamefont {Wang}}, \bibinfo {author} {\bibfnamefont
  {L.}~\bibnamefont {Chen}}, \ and\ \bibinfo {author} {\bibfnamefont {G.~J.}\
  \bibnamefont {Snyder}},\ }\bibfield  {title} {\enquote {\bibinfo {title}
  {Convergence of electronic bands for high performance bulk
  thermoelectrics},}\ }\href@noop {} {\bibfield  {journal} {\bibinfo  {journal}
  {Nature}\ }\textbf {\bibinfo {volume} {473}},\ \bibinfo {pages} {66--69}
  (\bibinfo {year} {2011})}\BibitemShut {NoStop}%
\bibitem [{\citenamefont {LaLonde}\ \emph {et~al.}(2011)\citenamefont
  {LaLonde}, \citenamefont {Pei}, \citenamefont {Wang},\ and\ \citenamefont
  {Snyder}}]{lalonde2011lead}%
  \BibitemOpen
  \bibfield  {author} {\bibinfo {author} {\bibfnamefont {A.~D.}\ \bibnamefont
  {LaLonde}}, \bibinfo {author} {\bibfnamefont {Y.}~\bibnamefont {Pei}},
  \bibinfo {author} {\bibfnamefont {H.}~\bibnamefont {Wang}}, \ and\ \bibinfo
  {author} {\bibfnamefont {G.~J.}\ \bibnamefont {Snyder}},\ }\bibfield  {title}
  {\enquote {\bibinfo {title} {Lead telluride alloy thermoelectrics},}\
  }\href@noop {} {\bibfield  {journal} {\bibinfo  {journal} {Materials today}\
  }\textbf {\bibinfo {volume} {14}},\ \bibinfo {pages} {526--532} (\bibinfo
  {year} {2011})}\BibitemShut {NoStop}%
\bibitem [{\citenamefont {Sootsman}, \citenamefont {Chung},\ and\ \citenamefont
  {Kanatzidis}(2009)}]{sootsman2009new}%
  \BibitemOpen
  \bibfield  {author} {\bibinfo {author} {\bibfnamefont {J.~R.}\ \bibnamefont
  {Sootsman}}, \bibinfo {author} {\bibfnamefont {D.~Y.}\ \bibnamefont {Chung}},
  \ and\ \bibinfo {author} {\bibfnamefont {M.~G.}\ \bibnamefont {Kanatzidis}},\
  }\bibfield  {title} {\enquote {\bibinfo {title} {New and old concepts in
  thermoelectric materials},}\ }\href@noop {} {\bibfield  {journal} {\bibinfo
  {journal} {Angewandte Chemie International Edition}\ }\textbf {\bibinfo
  {volume} {48}},\ \bibinfo {pages} {8616--8639} (\bibinfo {year}
  {2009})}\BibitemShut {NoStop}%
\bibitem [{\citenamefont {Felser}, \citenamefont {Fecher},\ and\ \citenamefont
  {Balke}(2007)}]{felser2007spintronics}%
  \BibitemOpen
  \bibfield  {author} {\bibinfo {author} {\bibfnamefont {C.}~\bibnamefont
  {Felser}}, \bibinfo {author} {\bibfnamefont {G.~H.}\ \bibnamefont {Fecher}},
  \ and\ \bibinfo {author} {\bibfnamefont {B.}~\bibnamefont {Balke}},\
  }\bibfield  {title} {\enquote {\bibinfo {title} {Spintronics: a challenge for
  materials science and solid-state chemistry},}\ }\href@noop {} {\bibfield
  {journal} {\bibinfo  {journal} {Angewandte Chemie International Edition}\
  }\textbf {\bibinfo {volume} {46}},\ \bibinfo {pages} {668--699} (\bibinfo
  {year} {2007})}\BibitemShut {NoStop}%
\bibitem [{\citenamefont {Heremans}\ \emph {et~al.}(2008)\citenamefont
  {Heremans}, \citenamefont {Jovovic}, \citenamefont {Toberer}, \citenamefont
  {Saramat}, \citenamefont {Kurosaki}, \citenamefont {Charoenphakdee},
  \citenamefont {Yamanaka},\ and\ \citenamefont
  {Snyder}}]{heremans2008enhancement}%
  \BibitemOpen
  \bibfield  {author} {\bibinfo {author} {\bibfnamefont {J.~P.}\ \bibnamefont
  {Heremans}}, \bibinfo {author} {\bibfnamefont {V.}~\bibnamefont {Jovovic}},
  \bibinfo {author} {\bibfnamefont {E.~S.}\ \bibnamefont {Toberer}}, \bibinfo
  {author} {\bibfnamefont {A.}~\bibnamefont {Saramat}}, \bibinfo {author}
  {\bibfnamefont {K.}~\bibnamefont {Kurosaki}}, \bibinfo {author}
  {\bibfnamefont {A.}~\bibnamefont {Charoenphakdee}}, \bibinfo {author}
  {\bibfnamefont {S.}~\bibnamefont {Yamanaka}}, \ and\ \bibinfo {author}
  {\bibfnamefont {G.~J.}\ \bibnamefont {Snyder}},\ }\bibfield  {title}
  {\enquote {\bibinfo {title} {Enhancement of thermoelectric efficiency in pbte
  by distortion of the electronic density of states},}\ }\href@noop {}
  {\bibfield  {journal} {\bibinfo  {journal} {Science}\ }\textbf {\bibinfo
  {volume} {321}},\ \bibinfo {pages} {554--557} (\bibinfo {year}
  {2008})}\BibitemShut {NoStop}%
\bibitem [{\citenamefont {Zhao}\ \emph {et~al.}(2014)\citenamefont {Zhao},
  \citenamefont {Lo}, \citenamefont {Zhang}, \citenamefont {Sun}, \citenamefont
  {Tan}, \citenamefont {Uher}, \citenamefont {Wolverton}, \citenamefont
  {Dravid},\ and\ \citenamefont {Kanatzidis}}]{zhao2014ultralow}%
  \BibitemOpen
  \bibfield  {author} {\bibinfo {author} {\bibfnamefont {L.-D.}\ \bibnamefont
  {Zhao}}, \bibinfo {author} {\bibfnamefont {S.-H.}\ \bibnamefont {Lo}},
  \bibinfo {author} {\bibfnamefont {Y.}~\bibnamefont {Zhang}}, \bibinfo
  {author} {\bibfnamefont {H.}~\bibnamefont {Sun}}, \bibinfo {author}
  {\bibfnamefont {G.}~\bibnamefont {Tan}}, \bibinfo {author} {\bibfnamefont
  {C.}~\bibnamefont {Uher}}, \bibinfo {author} {\bibfnamefont {C.}~\bibnamefont
  {Wolverton}}, \bibinfo {author} {\bibfnamefont {V.~P.}\ \bibnamefont
  {Dravid}}, \ and\ \bibinfo {author} {\bibfnamefont {M.~G.}\ \bibnamefont
  {Kanatzidis}},\ }\bibfield  {title} {\enquote {\bibinfo {title} {Ultralow
  thermal conductivity and high thermoelectric figure of merit in snse
  crystals},}\ }\href@noop {} {\bibfield  {journal} {\bibinfo  {journal}
  {Nature}\ }\textbf {\bibinfo {volume} {508}},\ \bibinfo {pages} {373--377}
  (\bibinfo {year} {2014})}\BibitemShut {NoStop}%
\bibitem [{\citenamefont {Biswas}\ \emph {et~al.}(2012)\citenamefont {Biswas},
  \citenamefont {He}, \citenamefont {Blum}, \citenamefont {Wu}, \citenamefont
  {Hogan}, \citenamefont {Seidman}, \citenamefont {Dravid},\ and\ \citenamefont
  {Kanatzidis}}]{biswas2012high}%
  \BibitemOpen
  \bibfield  {author} {\bibinfo {author} {\bibfnamefont {K.}~\bibnamefont
  {Biswas}}, \bibinfo {author} {\bibfnamefont {J.}~\bibnamefont {He}}, \bibinfo
  {author} {\bibfnamefont {I.~D.}\ \bibnamefont {Blum}}, \bibinfo {author}
  {\bibfnamefont {C.-I.}\ \bibnamefont {Wu}}, \bibinfo {author} {\bibfnamefont
  {T.~P.}\ \bibnamefont {Hogan}}, \bibinfo {author} {\bibfnamefont {D.~N.}\
  \bibnamefont {Seidman}}, \bibinfo {author} {\bibfnamefont {V.~P.}\
  \bibnamefont {Dravid}}, \ and\ \bibinfo {author} {\bibfnamefont {M.~G.}\
  \bibnamefont {Kanatzidis}},\ }\bibfield  {title} {\enquote {\bibinfo {title}
  {High-performance bulk thermoelectrics with all-scale hierarchical
  architectures},}\ }\href@noop {} {\bibfield  {journal} {\bibinfo  {journal}
  {Nature}\ }\textbf {\bibinfo {volume} {489}},\ \bibinfo {pages} {414--418}
  (\bibinfo {year} {2012})}\BibitemShut {NoStop}%
\bibitem [{\citenamefont {Toberer}, \citenamefont {Zevalkink},\ and\
  \citenamefont {Snyder}(2011)}]{toberer2011phonon}%
  \BibitemOpen
  \bibfield  {author} {\bibinfo {author} {\bibfnamefont {E.~S.}\ \bibnamefont
  {Toberer}}, \bibinfo {author} {\bibfnamefont {A.}~\bibnamefont {Zevalkink}},
  \ and\ \bibinfo {author} {\bibfnamefont {G.~J.}\ \bibnamefont {Snyder}},\
  }\bibfield  {title} {\enquote {\bibinfo {title} {Phonon engineering through
  crystal chemistry},}\ }\href@noop {} {\bibfield  {journal} {\bibinfo
  {journal} {Journal of Materials Chemistry}\ }\textbf {\bibinfo {volume}
  {21}},\ \bibinfo {pages} {15843--15852} (\bibinfo {year} {2011})}\BibitemShut
  {NoStop}%
\bibitem [{\citenamefont {Ma}\ \emph {et~al.}(2017)\citenamefont {Ma},
  \citenamefont {Hegde}, \citenamefont {Munira}, \citenamefont {Xie},
  \citenamefont {Keshavarz}, \citenamefont {Mildebrath}, \citenamefont
  {Wolverton}, \citenamefont {Ghosh},\ and\ \citenamefont
  {Butler}}]{ma2017computational}%
  \BibitemOpen
  \bibfield  {author} {\bibinfo {author} {\bibfnamefont {J.}~\bibnamefont
  {Ma}}, \bibinfo {author} {\bibfnamefont {V.~I.}\ \bibnamefont {Hegde}},
  \bibinfo {author} {\bibfnamefont {K.}~\bibnamefont {Munira}}, \bibinfo
  {author} {\bibfnamefont {Y.}~\bibnamefont {Xie}}, \bibinfo {author}
  {\bibfnamefont {S.}~\bibnamefont {Keshavarz}}, \bibinfo {author}
  {\bibfnamefont {D.~T.}\ \bibnamefont {Mildebrath}}, \bibinfo {author}
  {\bibfnamefont {C.}~\bibnamefont {Wolverton}}, \bibinfo {author}
  {\bibfnamefont {A.~W.}\ \bibnamefont {Ghosh}}, \ and\ \bibinfo {author}
  {\bibfnamefont {W.}~\bibnamefont {Butler}},\ }\bibfield  {title} {\enquote
  {\bibinfo {title} {Computational investigation of half-heusler compounds for
  spintronics applications},}\ }\href@noop {} {\bibfield  {journal} {\bibinfo
  {journal} {Physical Review B}\ }\textbf {\bibinfo {volume} {95}},\ \bibinfo
  {pages} {024411} (\bibinfo {year} {2017})}\BibitemShut {NoStop}%
\bibitem [{\citenamefont {Kandpal}, \citenamefont {Fecher},\ and\ \citenamefont
  {Felser}(2007)}]{kandpal2007calculated}%
  \BibitemOpen
  \bibfield  {author} {\bibinfo {author} {\bibfnamefont {H.~C.}\ \bibnamefont
  {Kandpal}}, \bibinfo {author} {\bibfnamefont {G.~H.}\ \bibnamefont {Fecher}},
  \ and\ \bibinfo {author} {\bibfnamefont {C.}~\bibnamefont {Felser}},\
  }\bibfield  {title} {\enquote {\bibinfo {title} {Calculated electronic and
  magnetic properties of the half-metallic, transition metal based heusler
  compounds},}\ }\href@noop {} {\bibfield  {journal} {\bibinfo  {journal}
  {Journal of Physics D: Applied Physics}\ }\textbf {\bibinfo {volume} {40}},\
  \bibinfo {pages} {1507} (\bibinfo {year} {2007})}\BibitemShut {NoStop}%
\bibitem [{\citenamefont {Blaha}\ \emph {et~al.}(2001)\citenamefont {Blaha},
  \citenamefont {Schwarz}, \citenamefont {Madsen}, \citenamefont {Kvasnicka},
  \citenamefont {Luitz} \emph {et~al.}}]{blaha2001wien2k}%
  \BibitemOpen
  \bibfield  {author} {\bibinfo {author} {\bibfnamefont {P.}~\bibnamefont
  {Blaha}}, \bibinfo {author} {\bibfnamefont {K.}~\bibnamefont {Schwarz}},
  \bibinfo {author} {\bibfnamefont {G.~K.}\ \bibnamefont {Madsen}}, \bibinfo
  {author} {\bibfnamefont {D.}~\bibnamefont {Kvasnicka}}, \bibinfo {author}
  {\bibfnamefont {J.}~\bibnamefont {Luitz}},  \emph {et~al.},\ }\bibfield
  {title} {\enquote {\bibinfo {title} {wien2k},}\ }\href@noop {} {\bibfield
  {journal} {\bibinfo  {journal} {An augmented plane wave+ local orbitals
  program for calculating crystal properties}\ } (\bibinfo {year}
  {2001})}\BibitemShut {NoStop}%
\bibitem [{\citenamefont {Giannozzi}\ \emph {et~al.}(2009)\citenamefont
  {Giannozzi}, \citenamefont {Baroni}, \citenamefont {Bonini}, \citenamefont
  {Calandra}, \citenamefont {Car}, \citenamefont {Cavazzoni}, \citenamefont
  {Ceresoli}, \citenamefont {Chiarotti}, \citenamefont {Cococcioni},
  \citenamefont {Dabo} \emph {et~al.}}]{giannozzi2009quantum}%
  \BibitemOpen
  \bibfield  {author} {\bibinfo {author} {\bibfnamefont {P.}~\bibnamefont
  {Giannozzi}}, \bibinfo {author} {\bibfnamefont {S.}~\bibnamefont {Baroni}},
  \bibinfo {author} {\bibfnamefont {N.}~\bibnamefont {Bonini}}, \bibinfo
  {author} {\bibfnamefont {M.}~\bibnamefont {Calandra}}, \bibinfo {author}
  {\bibfnamefont {R.}~\bibnamefont {Car}}, \bibinfo {author} {\bibfnamefont
  {C.}~\bibnamefont {Cavazzoni}}, \bibinfo {author} {\bibfnamefont
  {D.}~\bibnamefont {Ceresoli}}, \bibinfo {author} {\bibfnamefont {G.~L.}\
  \bibnamefont {Chiarotti}}, \bibinfo {author} {\bibfnamefont {M.}~\bibnamefont
  {Cococcioni}}, \bibinfo {author} {\bibfnamefont {I.}~\bibnamefont {Dabo}},
  \emph {et~al.},\ }\bibfield  {title} {\enquote {\bibinfo {title} {Quantum
  espresso: a modular and open-source software project for quantum simulations
  of materials},}\ }\href@noop {} {\bibfield  {journal} {\bibinfo  {journal}
  {Journal of physics: Condensed matter}\ }\textbf {\bibinfo {volume} {21}},\
  \bibinfo {pages} {395502} (\bibinfo {year} {2009})}\BibitemShut {NoStop}%
\bibitem [{\citenamefont {Perdew}, \citenamefont {Burke},\ and\ \citenamefont
  {Ernzerhof}(1996)}]{perdew1996generalized}%
  \BibitemOpen
  \bibfield  {author} {\bibinfo {author} {\bibfnamefont {J.~P.}\ \bibnamefont
  {Perdew}}, \bibinfo {author} {\bibfnamefont {K.}~\bibnamefont {Burke}}, \
  and\ \bibinfo {author} {\bibfnamefont {M.}~\bibnamefont {Ernzerhof}},\
  }\bibfield  {title} {\enquote {\bibinfo {title} {Generalized gradient
  approximation made simple},}\ }\href@noop {} {\bibfield  {journal} {\bibinfo
  {journal} {Physical review letters}\ }\textbf {\bibinfo {volume} {77}},\
  \bibinfo {pages} {3865} (\bibinfo {year} {1996})}\BibitemShut {NoStop}%
\bibitem [{\citenamefont {Tran}\ and\ \citenamefont
  {Blaha}(2009)}]{PhysRevLett.102.226401}%
  \BibitemOpen
  \bibfield  {author} {\bibinfo {author} {\bibfnamefont {F.}~\bibnamefont
  {Tran}}\ and\ \bibinfo {author} {\bibfnamefont {P.}~\bibnamefont {Blaha}},\
  }\bibfield  {title} {\enquote {\bibinfo {title} {Accurate band gaps of
  semiconductors and insulators with a semilocal exchange-correlation
  potential},}\ }\href {\doibase 10.1103/PhysRevLett.102.226401} {\bibfield
  {journal} {\bibinfo  {journal} {Phys. Rev. Lett.}\ }\textbf {\bibinfo
  {volume} {102}},\ \bibinfo {pages} {226401} (\bibinfo {year}
  {2009})}\BibitemShut {NoStop}%
\bibitem [{\citenamefont {Madsen}\ and\ \citenamefont
  {Singh}(2006)}]{madsen2006boltztrap}%
  \BibitemOpen
  \bibfield  {author} {\bibinfo {author} {\bibfnamefont {G.~K.}\ \bibnamefont
  {Madsen}}\ and\ \bibinfo {author} {\bibfnamefont {D.~J.}\ \bibnamefont
  {Singh}},\ }\bibfield  {title} {\enquote {\bibinfo {title} {Boltztrap. a code
  for calculating band-structure dependent quantities},}\ }\href@noop {}
  {\bibfield  {journal} {\bibinfo  {journal} {Computer Physics Communications}\
  }\textbf {\bibinfo {volume} {175}},\ \bibinfo {pages} {67--71} (\bibinfo
  {year} {2006})}\BibitemShut {NoStop}%
\bibitem [{\citenamefont {Togo}\ and\ \citenamefont
  {Tanaka}(2015)}]{togo2015first}%
  \BibitemOpen
  \bibfield  {author} {\bibinfo {author} {\bibfnamefont {A.}~\bibnamefont
  {Togo}}\ and\ \bibinfo {author} {\bibfnamefont {I.}~\bibnamefont {Tanaka}},\
  }\bibfield  {title} {\enquote {\bibinfo {title} {First principles phonon
  calculations in materials science},}\ }\href@noop {} {\bibfield  {journal}
  {\bibinfo  {journal} {Scripta Materialia}\ }\textbf {\bibinfo {volume}
  {108}},\ \bibinfo {pages} {1--5} (\bibinfo {year} {2015})}\BibitemShut
  {NoStop}%
\bibitem [{\citenamefont {Kumarasinghe}\ and\ \citenamefont
  {Neophytou}(2019)}]{kumarasinghe2019band}%
  \BibitemOpen
  \bibfield  {author} {\bibinfo {author} {\bibfnamefont {C.}~\bibnamefont
  {Kumarasinghe}}\ and\ \bibinfo {author} {\bibfnamefont {N.}~\bibnamefont
  {Neophytou}},\ }\bibfield  {title} {\enquote {\bibinfo {title} {Band
  alignment and scattering considerations for enhancing the thermoelectric
  power factor of complex materials: The case of co-based half-heusler
  alloys},}\ }\href@noop {} {\bibfield  {journal} {\bibinfo  {journal}
  {Physical Review B}\ }\textbf {\bibinfo {volume} {99}},\ \bibinfo {pages}
  {195202} (\bibinfo {year} {2019})}\BibitemShut {NoStop}%
\bibitem [{\citenamefont {Zhang}, \citenamefont {Du},\ and\ \citenamefont
  {Singh}(2010)}]{zhang2010zintl}%
  \BibitemOpen
  \bibfield  {author} {\bibinfo {author} {\bibfnamefont {L.}~\bibnamefont
  {Zhang}}, \bibinfo {author} {\bibfnamefont {M.-H.}\ \bibnamefont {Du}}, \
  and\ \bibinfo {author} {\bibfnamefont {D.~J.}\ \bibnamefont {Singh}},\
  }\bibfield  {title} {\enquote {\bibinfo {title} {Zintl-phase compounds with
  snsb 4 tetrahedral anions: Electronic structure and thermoelectric
  properties},}\ }\href@noop {} {\bibfield  {journal} {\bibinfo  {journal}
  {Physical Review B}\ }\textbf {\bibinfo {volume} {81}},\ \bibinfo {pages}
  {075117} (\bibinfo {year} {2010})}\BibitemShut {NoStop}%
\bibitem [{\citenamefont {Fu}\ \emph {et~al.}(2016)\citenamefont {Fu},
  \citenamefont {Wu}, \citenamefont {Liu}, \citenamefont {He}, \citenamefont
  {Zhao},\ and\ \citenamefont {Zhu}}]{fu2016enhancing}%
  \BibitemOpen
  \bibfield  {author} {\bibinfo {author} {\bibfnamefont {C.}~\bibnamefont
  {Fu}}, \bibinfo {author} {\bibfnamefont {H.}~\bibnamefont {Wu}}, \bibinfo
  {author} {\bibfnamefont {Y.}~\bibnamefont {Liu}}, \bibinfo {author}
  {\bibfnamefont {J.}~\bibnamefont {He}}, \bibinfo {author} {\bibfnamefont
  {X.}~\bibnamefont {Zhao}}, \ and\ \bibinfo {author} {\bibfnamefont
  {T.}~\bibnamefont {Zhu}},\ }\bibfield  {title} {\enquote {\bibinfo {title}
  {Enhancing the figure of merit of heavy-band thermoelectric materials through
  hierarchical phonon scattering},}\ }\href@noop {} {\bibfield  {journal}
  {\bibinfo  {journal} {Advanced Science}\ }\textbf {\bibinfo {volume} {3}},\
  \bibinfo {pages} {1600035} (\bibinfo {year} {2016})}\BibitemShut {NoStop}%
\bibitem [{\citenamefont {Hori}\ \emph {et~al.}(2020)\citenamefont {Hori},
  \citenamefont {Minami}, \citenamefont {Saito},\ and\ \citenamefont
  {Ishii}}]{hori2020first}%
  \BibitemOpen
  \bibfield  {author} {\bibinfo {author} {\bibfnamefont {A.}~\bibnamefont
  {Hori}}, \bibinfo {author} {\bibfnamefont {S.}~\bibnamefont {Minami}},
  \bibinfo {author} {\bibfnamefont {M.}~\bibnamefont {Saito}}, \ and\ \bibinfo
  {author} {\bibfnamefont {F.}~\bibnamefont {Ishii}},\ }\bibfield  {title}
  {\enquote {\bibinfo {title} {First-principles calculation of lattice thermal
  conductivity and thermoelectric figure of merit in ferromagnetic half-heusler
  alloy comnsb},}\ }\href@noop {} {\bibfield  {journal} {\bibinfo  {journal}
  {Applied Physics Letters}\ }\textbf {\bibinfo {volume} {116}},\ \bibinfo
  {pages} {242408} (\bibinfo {year} {2020})}\BibitemShut {NoStop}%
\bibitem [{\citenamefont {Mishra}, \citenamefont {Satpathy},\ and\
  \citenamefont {Jepsen}(1997)}]{mishra1997electronic}%
  \BibitemOpen
  \bibfield  {author} {\bibinfo {author} {\bibfnamefont {S.}~\bibnamefont
  {Mishra}}, \bibinfo {author} {\bibfnamefont {S.}~\bibnamefont {Satpathy}}, \
  and\ \bibinfo {author} {\bibfnamefont {O.}~\bibnamefont {Jepsen}},\
  }\bibfield  {title} {\enquote {\bibinfo {title} {Electronic structure and
  thermoelectric properties of bismuth telluride and bismuth selenide},}\
  }\href@noop {} {\bibfield  {journal} {\bibinfo  {journal} {Journal of
  Physics: Condensed Matter}\ }\textbf {\bibinfo {volume} {9}},\ \bibinfo
  {pages} {461} (\bibinfo {year} {1997})}\BibitemShut {NoStop}%
\bibitem [{\citenamefont {Min}\ \emph {et~al.}(2013)\citenamefont {Min},
  \citenamefont {Roh}, \citenamefont {Yang}, \citenamefont {Park},
  \citenamefont {Kim}, \citenamefont {Hwang}, \citenamefont {Lee},
  \citenamefont {Lee},\ and\ \citenamefont {Jeong}}]{min2013surfactant}%
  \BibitemOpen
  \bibfield  {author} {\bibinfo {author} {\bibfnamefont {Y.}~\bibnamefont
  {Min}}, \bibinfo {author} {\bibfnamefont {J.~W.}\ \bibnamefont {Roh}},
  \bibinfo {author} {\bibfnamefont {H.}~\bibnamefont {Yang}}, \bibinfo {author}
  {\bibfnamefont {M.}~\bibnamefont {Park}}, \bibinfo {author} {\bibfnamefont
  {S.~I.}\ \bibnamefont {Kim}}, \bibinfo {author} {\bibfnamefont
  {S.}~\bibnamefont {Hwang}}, \bibinfo {author} {\bibfnamefont {S.~M.}\
  \bibnamefont {Lee}}, \bibinfo {author} {\bibfnamefont {K.~H.}\ \bibnamefont
  {Lee}}, \ and\ \bibinfo {author} {\bibfnamefont {U.}~\bibnamefont {Jeong}},\
  }\bibfield  {title} {\enquote {\bibinfo {title} {Surfactant-free scalable
  synthesis of bi2te3 and bi2se3 nanoflakes and enhanced thermoelectric
  properties of their nanocomposites},}\ }\href@noop {} {\bibfield  {journal}
  {\bibinfo  {journal} {Advanced Materials}\ }\textbf {\bibinfo {volume}
  {25}},\ \bibinfo {pages} {1425--1429} (\bibinfo {year} {2013})}\BibitemShut
  {NoStop}%
\bibitem [{\citenamefont {Lenoir}\ \emph {et~al.}(1996)\citenamefont {Lenoir},
  \citenamefont {Cassart}, \citenamefont {Michenaud}, \citenamefont
  {Scherrer},\ and\ \citenamefont {Scherrer}}]{lenoir1996transport}%
  \BibitemOpen
  \bibfield  {author} {\bibinfo {author} {\bibfnamefont {B.}~\bibnamefont
  {Lenoir}}, \bibinfo {author} {\bibfnamefont {M.}~\bibnamefont {Cassart}},
  \bibinfo {author} {\bibfnamefont {J.-P.}\ \bibnamefont {Michenaud}}, \bibinfo
  {author} {\bibfnamefont {H.}~\bibnamefont {Scherrer}}, \ and\ \bibinfo
  {author} {\bibfnamefont {S.}~\bibnamefont {Scherrer}},\ }\bibfield  {title}
  {\enquote {\bibinfo {title} {Transport properties of bi-rich bi-sb alloys},}\
  }\href@noop {} {\bibfield  {journal} {\bibinfo  {journal} {Journal of Physics
  and Chemistry of Solids}\ }\textbf {\bibinfo {volume} {57}},\ \bibinfo
  {pages} {89--99} (\bibinfo {year} {1996})}\BibitemShut {NoStop}%
\bibitem [{\citenamefont {Fang}\ \emph {et~al.}(2020)\citenamefont {Fang},
  \citenamefont {Li}, \citenamefont {Wu}, \citenamefont {Zhang}, \citenamefont
  {Zhao},\ and\ \citenamefont {Zhu}}]{fang2020anisotropic}%
  \BibitemOpen
  \bibfield  {author} {\bibinfo {author} {\bibfnamefont {T.}~\bibnamefont
  {Fang}}, \bibinfo {author} {\bibfnamefont {F.}~\bibnamefont {Li}}, \bibinfo
  {author} {\bibfnamefont {Y.}~\bibnamefont {Wu}}, \bibinfo {author}
  {\bibfnamefont {Q.}~\bibnamefont {Zhang}}, \bibinfo {author} {\bibfnamefont
  {X.}~\bibnamefont {Zhao}}, \ and\ \bibinfo {author} {\bibfnamefont
  {T.}~\bibnamefont {Zhu}},\ }\bibfield  {title} {\enquote {\bibinfo {title}
  {Anisotropic thermoelectric properties of n-type te-free (bi, sb) 2se3 with
  orthorhombic structure},}\ }\href@noop {} {\bibfield  {journal} {\bibinfo
  {journal} {ACS Applied Energy Materials}\ }\textbf {\bibinfo {volume} {3}},\
  \bibinfo {pages} {2070--2077} (\bibinfo {year} {2020})}\BibitemShut {NoStop}%
\end{thebibliography}%

\end{document}